\documentclass[12pt]{article}

\usepackage{smile}

\usepackage{natbib}
\usepackage{multirow}

\def\given{\,|\,}

\def\tr{\mathop{\text{tr}}\kern.2ex}

\providecommand{\norm}[1]{\vvvert#1\vvvert}

\makeatletter
\def\@biblabel#1{\hspace*{-\labelsep}}
\makeatother
\begin{document}

\title{\Huge An Overview on the Estimation of Large Covariance and Precision Matrices}

\author{Jianqing Fan\thanks{Address: Department of ORFE, Sherrerd Hall, Princeton University, Princeton, NJ 08544, USA, e-mail: \textit{jqfan@princeton.edu},
\textit{yuanliao@umd.edu}, \textit{hanliu@princeton.edu}. Fan's research was supported by National Institutes of Health grants R01-GM072611 and R01GM100474-01 and National Science Foundation grants DMS-1206464 and DMS-1308566.}, Yuan Liao$^\ddag$ and Han Liu$^*$
\medskip\\{\normalsize $^*$Department of Operations Research and Financial Engineering,  Princeton University}
\medskip\\{\normalsize $^\ddag$ Department of Mathematics, University of Maryland}}

\date{\today}

\maketitle

\begin{abstract}
Estimating large covariance and precision matrices are  fundamental in modern multivariate analysis. The problems arise from statistical analysis of large panel economics and finance data. The covariance matrix reveals  marginal correlations between  variables, while the precision matrix encodes conditional correlations between pairs of variables given the remaining variables.      In this paper, we provide a selective review of  several recent developments on estimating large covariance and precision matrices. We focus on two general approaches: rank based method and factor model based method. Theories and applications of both approaches are presented.  These methods are expected to be widely applicable  to analysis of economic and financial data.

\end{abstract}


\section{Introduction}

Estimating large covariance and  precision (inverse covariance) matrices   becomes fundamental problems in modern multivariate analysis, which find applications in many fields, ranging from economics and finance to biology, social networks, and health sciences \citep{fan2014challenges}.  When the dimension of the covariance matrix is large, the estimation problem is generally challenging. It is well-known that the sample covariance based on the observed data is singular when the dimension is larger than the sample size. In addition, the   aggregation of massive amount of estimation errors can make considerable adverse impacts on the estimation accuracy. Therefore, estimating large covariance and precision matrices  attracts rapidly growing research attentions in the past decade.

In recent years researchers have proposed various regularization techniques to consistently estimate large covariance and precision matrices. To estimate large covariance matrices, one of the key assumptions made in the literature is that the  target matrix of interest is sparse, namely,   many entries are zero or nearly so \citep{Bickel08a, lam2009sparsistency, el2010high,  rigollet2012estimation}.
To estimate large precision matrices, it is often the case that the precision matrix is sparse.  A commonly used method for estimating the sparse precision matrix is to employ an  $\ell_{1}$-penalized maximum likelihood, see  for instance, \cite{banerjee2008model,Yuan07, friedman2008sparse, rothman2008sparse}. To further reduce the estimation bias, \cite{lam2009sparsistency, Shen12} proposed non-convex penalties for sparse precision matrix estimation and studied their theoretical properties. For more general theory on penalized likelihood methods, see  \cite{fan2001variable, fan2004nonconcave, Zou06,Zhao06, bickel2009simultaneous, Wainwright09}.



The literature has  been further expanded into robust estimation based on  regularized rank-based approaches \citep{ liu2012high, xue2012regularized}.   The rank-based method is particularly appealing when the distribution of the data generating process is non-Gaussian and heavy-tailed. It is particularly appealing for analysis of financial data.  The literature includes, for instance, \cite{han2013optimal, wegkamp2013adaptive, mitra2014multivariate}, etc.  The heavy-tailed data are often modeled by the elliptical distribution family,  which has been widely used for financial data analysis. See \cite{owen1983class, hamada2004capm} and \cite{ frahm2008tyler}.

In addition, in many applications the sparsity property   is not directly applicable.  For example, financial returns depend on the equity market risks, housing prices depend on the economic health, gene expressions can be stimulated by cytokines, among others.
Due to the presence of common factors, it is unrealistic to assume that many outcomes are uncorrelated.
A natural extension is the \textit{conditional sparsity}, namely, conditional on the   common factors, the covariance matrix of the remaining components of the outcome variables is sparse. In order to do so, we consider a factor model.
The factor model is one of the most useful tools for understanding the common dependence among multivariate outputs, which  has broad applications in the  statistics and econometrics literature.  For instance, it is commonly used to measure the vector of economic outputs   or excessive returns of financial assets over time, and has been found to produce good  out-of-sample forecast for macroeconomic variables \citep{boivin2005understanding, stock2002forecasting}. In high dimensions,
 the unknown factors and loadings are typically estimated by the principal components  method, and the separations between the common factors and idiosyncratic components are characterized via \textit{pervasiveness} assumptions.  See, for instance, \cite{stock2002forecasting, bai2003inferential, BN02, fan2008high, BT11, onatski2012asymptotics, LamYao, POET}, among others.   In the statistical literature, the separations between the common factors and idiosyncratic components are carried out by the low-rank plus sparsity decomposition.  See, for example, \cite{candes2009exact,    koltchinskii2011nuclear, FLM11, negahban2011estimation, cai2013sparse, ma2013sparse}.

 In this paper, we provide a selective review of  several recent developments on estimating large covariance and precision matrices. We focus on two general approaches: rank-based method and factor model based method. 
 Theories and applications of both approaches are presented.  Note that this paper is not an exhaustive survey, and many other regularization methods are also commonly used in the literature, e.g.,  the shrinkage method \citep{ledoit2003improved, ledoit2004well}.  We refer to  \cite{fan2013statistical},  \cite{pourahmadi2013high} and the references therein for reviews of other commonly used methods.

This paper is organized as follows. Section 2 presents  methods of estimating sparse covariance matrices. Section 3 reviews methods of estimating sparse precision matrices.   Section 4 discusses robust covariance and precision matrix estimations using rank-based estimators. Sections 5 and 6 respectively presents factor models based method, respectively in the cases of observable and unobservable factors. Section 7 introduces the structured factor model.  Finally, Section 8 provides further discussions.

Let  $\lambda_{\min}(\Ab)$ and $\lambda_{\max}(\Ab)$ respectively denote the minimum and maximum eigenvalues of $\Ab$. Let $\psi_{\max}(\Ab)$ be the largest singular value of $\Ab$.
We shall use $\|\Ab\|_2$ and $\|\Ab\|_{\text{F}}$ to denote the operator norm and Frobenius norm of a matrix $\Ab$,  respectively defined as $\lambda^{1/2}_{\max}(\Ab'\Ab)$ and $\tr^{1/2}(\Ab'\Ab)$. Throughout this paper, we shall use $p$ and $T$  to respectively denote the dimension of the covariance matrix of interest, and the sample size. Let $\vb = (v_{1}, \ldots, v_{p})'\in \RR^{p}$ be a real valued vector, we define the vector norms: $\norm{\vb}_1 = \sum_{j=1}^{p}|v_{j}|,~\norm{\vb}_2^2 = \sum_{j=1}^{p}v_{j}^2,~\norm{\vb}_{\infty} = \max_{1 \leq j \leq p}|v_j|$. Let $\cS$ be a subspace of $\RR^{p}$, we use $\vb_{\cS}$ to denote the projection of $\vb$ onto $\cS$: $\vb_{\cS} = \argmin_{\ub \in \cS} \norm{\ub - \vb}_2^2$. We also define the orthogonal complement of $\cS$ as $\cS^\perp = \big\{ \ub \in \RR^{p}~\Big|~\ub'\vb=0,~\textrm{for~any}~\vb \in \cS\big\}$.
Let $\Ab\in\mathbb{R}^{p\times p}$ and $I, J\subset \{1,\ldots, N\}$ be two sets. Denote by $\Ab_{I,J}$  the submatrix of $\Ab$ with rows and columns indexed by $I$ and $J$.  Letting $\Ab_{*j} = (\Ab_{1j},...,\Ab_{pj})'$ and $\Ab_{k*} = (\Ab_{k1},...,\Ab_{kp})'$ denote the $j^{\rm th}$ column and $k^{\rm th}$ row of $\Ab$ in vector forms, we define the matrix norms: $\norm{\Ab}_1 = \max_{j}\norm{\Ab_{*j}}_1,~\norm{\Ab}_\infty = \max_{k}\norm{\Ab_{k*}}_1,~\norm{\Ab}_{\max} = \max_{j}\norm{\Ab_{*j}}_{\infty}$.    We also define matrix  elementwise (pseudo-) norms: $\norm{\Ab}_{1,\mathrm{off}} = \sum_{j \neq k}|\Ab_{jk}|$ and $\norm{\Ab}_{\infty,\mathrm{off}} = \max_{j \neq k}|\Ab_{jk}|$.  We write $a_{n} \asymp b_{n}$ if there are positive constants $c_{1}$ and $c_{2}$ independent of $n$ such that $c_{1}b_{n}\leq a_{n}\leq c_{2}b_{n}$.


\section{Estimating sparse covariance matrix }

Let $Y_{it}$ be the observed data  for the $i^{th}$ ($i=1,...,p$) individual at time $t=1,...,T$ (or the $t^{th}$ observation for the $i^{th}$ variable). We are interested in estimating the $p\times p$ covariance matrix  $\bSigma=(\sigma_{ij})_{p\times p}$ of  $\bY_{t}=(Y_{1t},...,Y_{pt})'$, assumed to be independent of $t$. The sample covariance matrix is defined as
$$
\Sbb=\frac{1}{T-1}\sum_{t=1}^T(\bY_t-\bar \bY)(\bY_t-\bar \bY)',\quad \bar \bY=\frac{1}{T}\sum_{t=1}^T\bY_t.
$$
When $p>T$, however, it is well-known that $\Sbb$ is singular. It also accumulates many estimation errors due to the large number of free parameters to estimate.

Sparsity is one of the most essential assumptions for high-dimensional covariance matrix estimation, which assumes that a majority of the off-diagonal elements are nearly zero, and effectively reduces the number of free parameters to estimate.  Specifically, it assumes that there is $q\geq 0$, so that the following defined quantity
\begin{equation}
m_p=\begin{cases}\label{e1}
\max_{i\leq p}\sum_{j=1}^p1\{\sigma_{ij}\neq0\},&\text{ if } q=0\\
\max_{i\leq p}\sum_{j=1}^p|\sigma_{ij}|^{q},&\text{ if } 0<q<1
\end{cases}
\end{equation}
is either bounded or grow slowly as $p\to\infty$. Here  $1\{\cdot\}$ denotes the indicator function.
Such an assumption is reasonable in many applications.  For instance, in a longitudinal study where variables have a natural order, variables are likely weakly correlated when they are far  apart \citep{wu2003nonparametric}.  Under the sparsity assumption, many regularization based estimation methods have been proposed.  This section selectively overviews several state-of-the-art statistical methods for estimating large sparse covariance matrices.

\subsection{Thresholding estimation}

One of the most convenient methods to estimate sparse covariance matrices  is the thresholding, which sets small estimated elements to zero \citep{Bickel08a}. Let $s_{ij}$ be the $(i,j)^{th}$ element of $\Sbb$. For a pre-specified thresholding value $\omega_T$, define
\begin{equation}\label{e2}
\widehat\bSigma=(\widehat\sigma_{ij})_{p\times p},\quad \widehat\sigma_{ij}=\begin{cases}s_{ij},& \text{ if } i=j\\
s_{ij}1\{|s_{ij}|>\omega_T\},&\text{ if } i\neq j
\end{cases}.
\end{equation} The thresholding value should dominate the maximum estimation error $\max_{i\neq j}|s_{ij}-\sigma_{ij}|$. When the data are Gaussian or sub-Gaussian, it can be taken as $$
\omega_T=C\sqrt{\frac{\log p}{T}}, \quad \text{ for some } C>0
$$
so that  the probability of the exception event $\{\max_{i\neq j}|s_{ij}-\sigma_{ij}| > \omega_T\}$ tends to zero very fast.

The advantage of thresholding is that it avoids estimating small elements so that noise does not accumulate. The decision of whether an element should be estimated is much easier than the attempt to estimate it accurately.  Indeed, under some regularity conditions, \cite{Bickel08a} showed that, if $m_p\omega_T^{1-q}\to0$ as $p,T\rightarrow \infty$,  we have
\begin{equation}\label{e3new}
\|\widehat\bSigma-\bSigma\|_2=O_P(m_p\omega_T^{1-q})~~\text{and}~~\|\widehat\bSigma^{-1}-\bSigma^{-1}\|_2=O_P(m_p\omega_T^{1-q}),
\end{equation}
where $m_p$ and $q$ are as defined in (\ref{e1}). In the case that all the ``small" elements of $\bSigma$ are exactly zero so that we take $q=0$, the above convergence rate becomes $O_P(\sqrt{\frac{\log p}{T}})$ if $m_p$ is bounded.  Since each element in the covariance matrix can be estimated with an error of order $O_P(T^{-1/2})$, it hence only costs us a $\log(p)$ factor to learn the unknown locations of the non-zero elements.

\subsection{Adaptive thresholding and entry-dependent thresholding}
The simple thresholding (\ref{e2}) does not   take the varying scales of the marginal standard deviations into account. One way to account this is to threshold on the t-type statistics. For example, using the simple thresholding, we can define the  \textit{adaptive thresholding} estimator \citep{Cai11b}:
\begin{equation}\label{e3}
\widehat\bSigma=(\widehat\sigma_{ij})_{p\times p},\quad \widehat\sigma_{ij}=\begin{cases}s_{ij},& \text{ if } i=j\\
s_{ij}1\{|s_{ij}|/\text{SE}(s_{ij})>\omega_T\},&\text{ if } i\neq j
\end{cases},
\end{equation}
where SE$(s_{ij})$ is the estimated standard error of $s_{ij}$.

A simpler method to take the scale into account is to directly apply thresholding
on the correlation matrix. Let $\Rb=\diag(\Sbb)^{-1/2}\Sbb\diag(\Sbb)^{-1/2}=(r_{ij})_{p\times p}$ be the sample correlation matrix. We  then apply the simple thresholding on the off-diagonal elements of $\Rb$, and obtain the thresholded correlation matrix $\Rb^{\mathcal{T}}$. So the $(i,j)^{th}$ element of $\Rb^{\mathcal{T}}$ is $r_{ij}1\{|r_{ij}|>\omega_T\}$ when $i\neq j$, and one if $i=j$. Then the estimated covariance matrix is defined as
$$
\widehat\bSigma^*=\diag(\Sbb)^{1/2}\Rb^{\mathcal{T}}\diag(\Sbb)^{1/2}.
$$
In particular, when $\omega_T=0$, it is exactly the sample covariance matrix since no thresholding is employed, whereas when $\omega_T=1$, it is a diagonal matrix with marginal sample variances on its diagonal. This form is more appropriate than the simple thresholding since it is thresholded on the standardized scale. Moreover, $\widehat\bSigma^*$ is equivalent
to applying the \textit{entry dependent thresholding}
$$
\omega_{T,ij}=\sqrt{s_{ii}s_{jj}}\omega_T
$$
to the original sample covariance $\Sbb$.

\subsection{Generalized thresholding}

The  introduced thresholding estimators (\ref{e2}) and (\ref{e3}) are based on a simple thresholding rule, known as the  \textit{hard-thresholding}. In regression and wavelet shrinkage contexts (see, for example, \cite{donoho1995wavelet}), hard thresholding performs worse than some more flexible regularization methods, such as  the soft-thresholding and the smoothly clipped absolute deviation (SCAD) \citep{fan2001variable}, which combine thresholding with shrinkages.  The estimates resulting from such shrinkage typically are continuous functions of the maximum likelihood estimates (under Gaussianity), a desirable property that is not shared by the hard thresholding method.

Therefore, the \textit{generalized thresholding} rules of \cite{antoniadis2001regularization} can be applied to  estimating large covariance matrices.
The generalized thresholding rule depends on a thresholding parameter $\omega_T$ and a shrinkage function $h(\cdot; \omega_T):\mathbb{R}\to\mathbb{R}$, which satisfies
$$(i)\quad |h(z,\omega_T)|\leq |z|;\qquad
(ii) \quad h(z;\omega_T)=0 \text{ for } |z|\leq\omega_T;\qquad
(iii)\quad  |h(z;\omega_T)-z|\leq\omega_T.$$
There are a number of useful thresholding functions that are commonly used in the literature. For instance, the soft-thresholding takes $h(z;\omega_T)=$sgn$(z)(|z|-\omega_T)_+$, where $(x)_+=\max\{x,0\}$. Moreover, the SCAD thresholding is a compromise between hard and soft thresholding, whose amount of shrinkage decreases as $|z|$ increases and hence results in a nearly unbiased estimation.  Another example is the MCP thresholding, proposed by \cite{zhang2010nearly}.

 We can then define a generalized thresholding covariance estimator:
 \begin{equation} \label{e5}
\widehat\bSigma=(\widehat\sigma_{ij})_{p\times p},\quad \widehat\sigma_{ij}=\begin{cases}s_{ij},& \text{ if } i=j\\
h(s_{ij};\omega_T),&\text{ if } i\neq j
\end{cases}.
\end{equation}
Note that this admits  the hard-thresholding estimator (\ref{e2}) as a special case by taking $h(z;\omega_T)=z1\{|z|>\omega_T\}$. Both the adaptive thresholding and entry dependent thresholding can also be incorporated, by respectively setting $h(s_{ij}, \text{SE}(s_{ij})\omega_T)$ and $h(s_{ij},\sqrt{s_{ii}s_{jj}}\omega_T)$ on the $(i,j)^{th}$ element of the estimated covariance matrix when $i\neq j$. In addition, it is shown by  \cite{rothman2009generalized} that the use of generalized thresholding rules does not affect the rate of convergence in (\ref{e3new}), but it increases the family of shrinkages.

\subsection{Positive definiteness}

If the covariance matrix is sparse, it then follows from (\ref{e3new}) that the thresholding estimator $\widehat\bSigma$ is asymptotically  positive definite. On the other hand, it is often more desirable to require the positive definiteness under finite samples.  We discuss two approaches to achieving the finite sample positive definiteness.

\subsubsection{Choosing the thresholding constant}

\begin{figure}[htbp]
\begin{center}
\caption{Minimum eigenvalue of $\widehat\bSigma(C)$ as  a function of $C$ for three choices of thresholding rules. When the minimum eigenvalue reaches its maximum value, the covariance estimator becomes diagonal.}
\includegraphics[width=10cm]{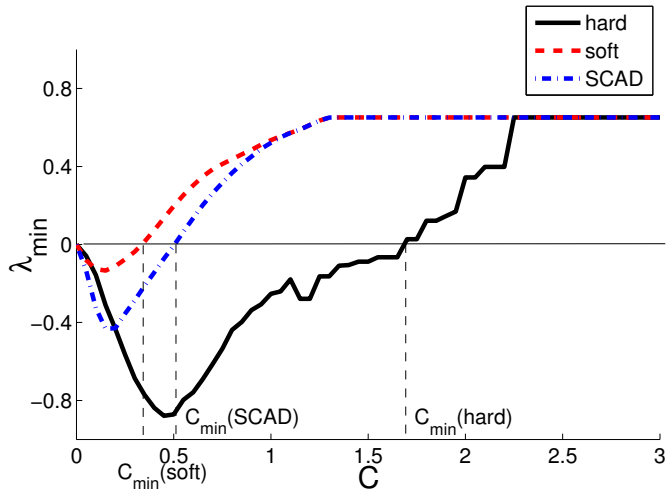}
\label{mineig}
\end{center}

\end{figure}

For simplicity, we focus on the  constant thresholding value $\omega_{T, ij}=\omega_T$; the case of entry-dependent thresholding  can be dealt similarly. The finite sample positive definiteness depends on the choice of the thresholding value $\omega_T$, which also depends on a prescribed constant $C$ through $\omega_T=C\sqrt{\frac{\log p}{T}}$. We   write $\widehat\bSigma(C)=\widehat\bSigma$ to emphasize its dependence on $C$.  When $C$ is sufficiently large, the estimator becomes diagonal, and its minimum eigenvalue is strictly positive.  We can then decreases the choice of $C$ until it reaches
$$
C_{\min}=\inf\{C>0: \lambda_{\min}(\widehat\bSigma(M))>0,\quad \forall M>C\}.
$$
 Thus, $C_{\min}$ is well defined and for all $C>C_{\min}$, $\widehat\bSigma(C)$ is positive definite under finite sample. We can obtain  $C_{\min}$  by solving $
\lambda_{\min}(\widehat\bSigma(C))=0, C\neq0.
$
Figure \ref{mineig} plots the minimum eigenvalue of $ \widehat\bSigma(C)$ as a function of $C$ for a random simple from a Gaussian distribution with $p>T$, using three different thresholding rules. It is clearly seen from the figure that there is a range of $C$ in which the covariance estimator is  both positive definite and non-diagonal.   In practice, we can choose $C$ in the range $(C_{\min}+\epsilon, M)$ for a small $\epsilon$ and large enough $M$ by, e.g., cross-validations. This method was suggested by \cite{POET} in a more complicated setting. Moreover, we also see from Figure \ref{mineig} that the hard-thresholding rule yields the narrowest range for the choice $C$ to give both positive definiteness and the non-diagonality.


\subsubsection{Nearest positive definite matrices}

An alternative approach to achieving the finite sample positive definiteness is through solving a constraint optimization problem. \cite{qi2006quadratically} introduced an algorithm for computing the \textit{nearest correlation matrix}: recall that $\Rb^{\mathcal{T}}$ is the thresholded correlation matrix, defined in  Section 2.2,  we find its nearest positive definite correlation matrix $\widehat\Rb$ by  solving:
$$
\widehat\Rb=\argmin_{\Ab}\|\Rb^{\mathcal{T}}-\Ab\|_\text{F}^2, \quad \text{s.t. } \Ab\geq0, \diag(\Ab)=\Ib_p.
$$
We can then transform back to the covariance  matrix as:
$
\widehat\bSigma=\diag(\Sbb)^{1/2}\widehat\Rb\diag(\Sbb)^{1/2}.$ Note that if  $\Rb^{\mathcal{T}}$ itself is positive semi-definite, $\widehat\Rb=\Rb^{\mathcal{T}}$; otherwise $\widehat\Rb$ is the nearest positive semi-definite correlation matrix. This  procedure is often called ``nearest correlation matrix projection", and can be solved effectively using the R-package   ``nearPD".

The  nearest correlation matrix projection, however, does not necessarily result in a sparse solution when $\Rb^{\mathcal{T}}$  is not positive definite.  \cite{liu2014sparse} introduced a covariance estimation method named EC2 (Estimation of Covariance with Eigenvalue Constraints). To motivate this method, note that the thresholding method  (\ref{e5}) can be equivalently casted as the solution to a penalized least squares problem:
$$
\widehat\bSigma=\argmin_{\Sigma=(\sigma_{ij})}\bigg{\{}\frac{1}{2}\|\Sbb-\bSigma\|_{\text{F}}^2+\sum_{i\neq j}P_{\omega_T}(\sigma_{ij})\bigg{\}}
$$
where $P_{\omega_T}(\cdot)$ is a penalty function, which corresponds to the shrinkage function $h(\cdot,\omega_T)$. For instance, when $$
P_{\omega_T}(t)=\omega_T^2-(|t|-\omega_T)^21\{|t|<\omega_T\},
$$
the solution is the hard-thresholding estimator (\ref{e2}) (\cite{antoniadis1997wavelets}). See \cite{antoniadis2001regularization} for the corresponding penalty functions of several popular shrinkage functions. The sparsity of the resulting estimator is hence due to the penalizations.  We can modify the above penalized least squares problem by adding an extra constraint to obtain positive definiteness:
 \begin{equation}\label{e6}
\widetilde\bSigma=\argmin_{\lambda_{\min}(\Sigma)\geq\tau}\bigg{\{}\frac{1}{2}\|\Sbb-\bSigma\|_{\text{F}}^2+\sum_{i\neq j}P_{\omega_T}(\sigma_{ij})\bigg{\}}
\end{equation}
where $\tau>0$ is a pre-specified tuning parameter that controls the smallest eigenvalue of the estimated covariance matrix $\widetilde\bSigma$. As a result, both sparsity and positive definiteness are guaranteed.  \cite{liu2014sparse}  showed that the problem (\ref{e6}) is convex when the penalty
function is convex,  and develops an  efficient  algorithm to solve it.  More details on the algorithm and theory of this estimator will be explained in   Section 4.


\section{Estimating sparse precision matrix}

Estimating a large inverse covariance matrix $\bTheta =\bSigma^{-1}$ is another fundamental problem in modern multivariate analysis. Unlike the covariance matrix $\bSigma $ which only captures the marginal correlations among  $\bY_t=(Y_{1t},\ldots, Y_{pt})'$, the inverse covariance matrix $\bTheta $ captures the conditional correlations among these variables and is closely related to undirected graphs under a Gaussian model.


More specifically,  we define an undirected graph  $G=(V,E)$, where $V$ contains nodes corresponding to the $p$ variables in $\bY_t$ and the edge  $(j,k)\in E$ if and only if  $\bTheta _{jk}\neq 0$.  Under a Gaussian model  $\bY_t\sim N(\bm{0}, {\bSigma} )$, the graph $G$ describes the conditional independence relationships among   $\bY_t=(Y_{1t},\ldots, Y_{pt})'$.  More specifically,  let $\bY_{t,\setminus \{j,k\}}=\left\{Y_{\ell t}:\ \ell\neq j,k\right\}$.    $Y_{jt}$ is independent of $Y_{kt}$ given $\bY_{t,\setminus \{j,k\}}$ for all $(j,k)\notin E$.

To illustrate the difference between the marginal and conditional uncorrelatedness.  We consider a Gaussian model $\bY_t \sim N(\bm{0}, \bSigma)$ with 
\begin{eqnarray*}
\bSigma =
\left(
\begin{array}{rrrrr}
1.05 & -0.23  &0.05 & -0.02& 0.05   \\
 -0.23 & 1.45  &-0.25 & 0.10& -0.25   \\
 0.05 & -0.25  &1.10 & -0.24 & 0.10   \\
  -0.02 & 0.10  &-0.24 & 1.10 & -0.24   \\
  0.05 & -0.25  &0.10 & -0.24 & 1.10  \\
\end{array}
\right)
~~\text{and}~~
\bTheta =
\left(
\begin{array}{rrrrr}
1 & 0.2 & 0 & 0 & 0   \\
 0.2 & 1  &0.2 & 0 & 0.2   \\
 0 & 0.2   &1 & 0.2 & 0   \\
  0 & 0  & 0.2 & 1  & 0.2   \\
  0 & 0.2  &0 & 0.2 & 1  \\
\end{array}
\right).
\end{eqnarray*}
We see that the inverse covariance matrix $\bTheta $ has many zero entries.  Thus the undirected graph $G$ defined by $\bTheta $ is sparse. However, the covariance matrix $\bSigma$ dense, which implies that every pair of  variables are marginally correlated. Thus the covariance matrix and inverse covariance matrix encode different relationships. For example, even though $Y_{1t}$ and $Y_{5t}$ are conditionally uncorrelated given the other variables, they are marginally correlated.   In addition to the graphical model problem, sparse precision matrix estimation has many other  applications. Examples  include high dimensional discriminant analysis \citep{cai2011constrained},  portfolio allocation \citep{fan2008high,fan2012vast}, principal component analysis,  and complex data visualization \citep{tokuda2011visualizing}.

Estimating the precision matrix $\bTheta$ requires very different techniques from estimating the covariance matrix. In the following subsections, we introduce several large precision estimation methods  under the assumption that $\bTheta$ is  sparse.  


\subsection{Penalized likelihood method}
One of the most commonly used approaches to estimating sparse precision matrices is through the maximum likelihood. When $\bY_1,\cdots,\bY_T$ are independently and identically distributed as $N(\bm{0}, {\bSigma} )$,  the negative Gaussian log-likelihood function is given by $\ell(\bTheta)=\tr(\Sbb\bTheta)-\log|\bTheta|$. When either the data are non-Gaussian or the data are weakly dependent, $\ell(\bTheta)$ becomes the quasi negative log-likelihood.  Nevertheless, we then consider the following penalized likelihood method:
$$
\widehat\bTheta=\argmin_{\bTheta=(\theta_{ij})_{p\times p}} \bigg{\{} \tr(\Sbb\bTheta)-\log|\bTheta| +\sum_{i\neq j}P_{\omega_T}(|\theta_{ij}|) \bigg{\}}
$$
where the penalty function $P_{\omega_T}(|\theta_{ij}|)$, defined the same way as in Section 2.4.2, encourages the sparsity of $\widehat\bTheta$.  One of the commonly used convex penalty is the $\ell_1$ penalty $P_{\omega_T}(t)=\omega_T|t|$, and the problem is then well studied in the literature (e.g., \cite{Yuan07, friedman2008sparse, banerjee2008model}).  Other related works are found in, e.g., \cite{meinshausen2006high, wille2004sparse}.

 In general, we recommend to use folded concave penalties such as SCAD and MCP, as these penalties do not introduce extra bias for estimating  nonzero entries with large absolute values \citep{lam2009sparsistency}.
Using local linear approximations, the penalized likelihood can be computed by an iterated reweighed Lasso: Given the estimate $\widehat\bTheta^{(k)}=(\widehat\theta_{ij}^{(k)})$ at the $k^{th}$ iteration, by the Taylor's expansion, we approximate
$$
P_{\omega_T}(|\theta_{ij}|)\approx P_{\omega_T}(|\widehat\theta_{ij}^{(k)}|)+
P_{\omega_T}'(|\widehat\theta_{ij}^{(k)}|)(|\theta_{ij}|-|\widehat\theta_{ij}^{(k)}|)
\equiv Q_{\omega_T}(|\theta_{ij}).
$$
The linear approximation $Q_{\omega_T}$ is the convex majorant of the  folded concave function at $|\widehat\theta_{ij}^{(k)}|$, namely, it satisfies
$$
P_{\omega_T}(|\theta_{ij}|) \leq Q_{\omega_T}(|\theta_{ij}|), \quad \mbox{and} \quad
P_{\omega_T}(|\widehat\theta_{ij}^{(k)}|) = Q_{\omega_T}(|\widehat\theta_{ij}^{(k)}|).
$$
Then the next iteration  is approximated by
\begin{equation}\label{e7}
\widehat\bTheta^{(k+1)}=\arg\min_{\bTheta=(\theta_{ij})}\bigg{\{}  \tr(\Sbb\bTheta)-\log|\bTheta| +\sum_{i\neq j} P_{\omega_T}'(|\widehat\theta_{ij}^{(k)}|)|\theta_{ij}|\bigg{\}}   +c,
\end{equation}
where $c$ is a constant that does not depend on $\bTheta$.  The problem (\ref{e7}) is convex and can be solved by the graphical Lasso algorithm of \cite{friedman2008sparse}.   Such an algorithm is called majorization-minimization algorithm \citep{lange2000optimization}. Since the penalty function is majorized from above,  it can easily be shown that the original objective function is decreasing in the iterations. Indeed, let
$
f(\bTheta) = \tr(\Sbb\bTheta)-\log|\bTheta| +\sum_{i\neq j}P_{\omega_T}(|\theta_{ij}|)
$
be the target value and $g(\bTheta)$ be its majorization function with $P_{\omega_T}(|\theta_{ij}|)$ replaced by $Q_{\omega_T}(|\theta_{ij}|)$.  Then,
$$
f(\widehat\bTheta^{(k+1)}) \leq g(\widehat\bTheta^{(k+1)})  \leq g(\widehat\bTheta^{(k)}) = f(\widehat\bTheta^{(k)}),
$$
where the first inequality follows from the majorization, the second inequality comes from the minimization, and the last equality follows the majorization at the point $\widehat\bTheta^{(k)}$.

Theoretical properties of $\widehat\bTheta$ have been thoroughly studied by \cite{rothman2008sparse} and \cite{ lam2009sparsistency}.

\subsection{Column-by-column estimation method}
\label{sec::columnregression}

Under the Gaussian model $\bY_t \sim N(\bm{0}, \bSigma)$, another approach  to estimating the precision matrix $\bTheta$ is through column-by-column regressions. For this,  \cite{Yuan10} and \cite{cai2011constrained}  propose the graphical Dantzig selector and CLIME respectively, which can be solved by linear programming.  More recently, \cite{Liu:arXiv1203.3896} and  \cite{Zhang12} propose the SCIO and scaled-Lasso methods. Compared to the penalized likelihood methods,  the column-by-column estimation methods are computationally simpler and   more amenable to theoretical analysis.

The  column-by-column precision matrix estimation method exploits the relationship between conditional distribution of multivariate Gaussian and linear regression. More specifically,  let $\bY\sim N(\bm{0}, \bSigma)$,  the conditional distribution of $Y_{j}$ given $\bY_{\setminus j}$ satisfies
\begin{eqnarray}
Y_j\given \bY_{\setminus j}\sim
N\bigl(\balpha_j'\bY_{\setminus j}\,,\, \sigma^{2}_{j}\bigr).  \nonumber
\end{eqnarray}
where
$\balpha_j=(\bSigma_{\setminus j,\setminus j})^{-1}\bSigma_{\setminus j,j}\in\mathbb{R}^{p-1}$ and $\sigma^{2}_{j} =\bSigma_{jj}-\bSigma_{\setminus j,  j}(\bSigma_{\setminus j,\setminus j})^{-1}\bSigma_{\setminus j,j}$. Hence, we can write
\begin{equation}
\label{gaussian.lin}
Y_j=\balpha_j' \bY_{\setminus j}+\epsilon_j,
\end{equation}
where $\epsilon_j\sim
N\bigl(0\,,\,
\sigma^{2}_{j}\bigr)$
is independent of $\bY_{\setminus j}$. Using the block matrix inversion formula, we have
\begin{equation}
\bTheta_{jj} = \sigma^{-2}_{j},\label{gaussian.sigma}, \qquad
\bTheta_{\setminus j,j}= -\sigma^{-2}_{j}\balpha_{j}.
\end{equation}
Therefore, we can recover $\bTheta$ in a column-by-column manner by regressing  $Y_j$ on $\bY_{\setminus j}$ for
$j=1,2,\cdots,p$.  For example, let $\Yb\in \mathbb{R}^{T\times p}$ be the data matrix. We denote by $\balpha_{j}:=(\alpha_{j1},\ldots, \alpha_{j(p-1)})'\in\mathbb{R}^{p-1}$. \cite{Meinshausen06} propose to estimate each $\balpha_{j}$ by solving the Lasso regression:
\begin{eqnarray}
\hat{\balpha}_{j} = \argmin_{\balpha_{j}\in\mathbb{R}^{p-1}}\frac{1}{2T}\bigl\|\Yb_{*j} - \Yb_{*\setminus j} \balpha_{j}\bigr\|^{2}_{2} + \lambda_{j}\bigl\|\balpha_{j}\bigr\|_{1},  \nonumber
\end{eqnarray}
where $\lambda_{j}$ is a tuning parameter.  Once  $\hat{\balpha}_{j}$ is obtained, we get the neighborhood edges by reading out the nonzero coefficients of $\balpha_{j}$. The final graph estimate $\hat{G}$ is obtained by either the ``AND'' or ``OR'' rule on combining the neighborhoods for all the $N$ nodes. To estimate $\bTheta$, we also estimate the $\sigma^2_j$'s  using the fitted sum of squared residuals  $\widehat{\sigma}^2_j=T^{-1}\bigl\|\Yb_{*j} - \Yb_{*\setminus j} \balpha_{j}\bigr\|^{2}_{2}$, then plug it into \eqref{gaussian.sigma}.

In another work, \cite{Yuan10} proposes to estimate $\balpha_{j}$ by solving the Dantzig selector:
\begin{eqnarray}
\hat{\balpha}_{j} = \argmin_{\balpha_{j}\in\mathbb{R}^{p-1}}\bigl\|\balpha_{j}\bigr\|_{1}~~\text{subject to}~~\bigl\|\Sbb_{\setminus j, j} - \Sbb_{\setminus j, \setminus j}\balpha_{j}\bigr\|_{\infty} \leq \gamma_{j},  \nonumber
\end{eqnarray}
where $\Sbb:=T^{-1}\Yb'\Yb$ is the sample covariance matrix and $\gamma_{j}$ is a tuning parameter. The constraint corresponds to a sample version of $\bSigma_{\setminus j,j} - \bSigma_{\setminus j,\setminus j} \balpha_j = 0$, with $\gamma_j$ indicating the estimation error.  Once $\hat{\balpha}_{j}$ is given, we can estimate $\sigma^{2}_{j}$ by $\hat{\sigma}^{2}_{j} = \bigl[1-2\hat{\balpha}'_{j}\Sbb_{\setminus j, j}  + \hat{\balpha}'_{j}\Sbb_{\setminus j, \setminus j} \hat{\balpha}_{j}\bigr]^{-1}$. We then obtain an estimator $\hat{\bTheta}$ of $\bTheta$ by plugging $\hat{\balpha}_{j}$ and $\hat{\sigma}^{2}_{j} $ into \eqref{gaussian.sigma}. \cite{Yuan10} analyzes the $L_{1}$-norm error $\|\hat{\bTheta} - \bTheta\|_{1}$ and shows its minimax optimality over certain model space.

More recently, \cite{Zhang12} propose to estimate $\balpha_{j}$ and $\sigma_{j}$ by solving a scaled-Lasso problem:
\begin{eqnarray}
\hat{\bbb}_{j}, \hat{\sigma}_{j} = \argmin_{\bbb=(b_{1},\ldots, b_{p})', \sigma}\biggl\{\frac{\bbb'\Sbb\bbb}{2\sigma}  + \frac{\sigma}{2} + \lambda\sum_{k=1}^{p} \Sbb_{kk}\bigl|b_{k}\bigr|~~\text{subject to}~b_{j}=-1\biggr\}.  \nonumber
\end{eqnarray}
Once $\hat{\bbb}_{j}$ is obtained, we estimate $\hat{\balpha}_{j}=(\hat{b}_1,\ldots, \hat{b}_{j-1}, \hat{b}_{j+1},\ldots, \hat{b}_p)'$ .  We then obtain the estimator  of $\bTheta$ by plugging $\hat{\balpha}_{j}$ and $\hat{\sigma}_{j} $ into\eqref{gaussian.sigma}.
\cite{Zhang12} provide the spectral-norm rate of convergence of the obtained precision matrix estimator.

Similar to the idea of the graphical Dantzig selector,  \cite{cai2011constrained} propose the CLIME estimator, which   stands
for ``Constrained $\ell_1$-Minimization for Inverse Matrix Estimation''. This method directly estimates the $j^{\rm th}$ column of  $\bTheta$ by solving
\begin{eqnarray}
\hat{\bTheta}_{*j} = \argmin_{\bTheta_{*j}}\bigl\|\bTheta_{*j} \bigr\|_{1}~~\text{subject to}~~\bigl\| \Sbb\bTheta_{*j} - \eb_{j}\bigr\|_{\infty}\leq \delta_{j},~~\text{for}~j=1,\ldots, p, \nonumber
\end{eqnarray}
where $\eb_{j}$ is the $j^{\rm th}$ canonical vector (i.e., the vector with the $j^{\rm th}$ element being 1, while the remaining elements being 0)  and $\delta_{j}$ is a tuning parameter. Again, the constraint represent a sample version of $\bSigma \bTheta_{*j} - \eb_{j} = 0$.  This optimization problem can be formulated into a linear program and has the potential to scale to large problems.  Under regularity conditions,  \cite{cai2011constrained}  show that  the estimator $ \widehat\bTheta$ is asymptotically positive definite, and derive its rate of convergence.

In a closely related work of CLIME, \cite{Liu:arXiv1203.3896} propose the SCIO estimator, which estimates the $j^{\rm th}$ column of $\bTheta$ by
\begin{eqnarray}
\hat{\bTheta}_{*j} = \argmin_{\bTheta_{*j}}\biggl\{ \frac{1}{2}\bTheta'_{*j}\Sbb\bTheta_{*j} - \eb'_{j}\bTheta_{*j}+\lambda_{j}\bigl\|\bTheta_{*j} \bigr\|_{1} \biggr\}. \nonumber
\end{eqnarray}
The SCIO estimator can be solved efficiently by the pathwise coordinate descent algorithm \citep{Friedman07}.

\subsection{ Tuning-insensitive precision matrix estimation}

Most of the methods described in the former subsection require choosing some tuning parameters  that control the
bias-variance tradeoff. Their theoretical justifications are  usually built on some theoretical choices of tuning
parameters that cannot be implemented in practice. For example, in the neighborhood pursuit method and the graphical Dantzig selector, the theoretically  optimal tuning parameters $\lambda_{j}$ and $\gamma_{j}$ depend on $\sigma^{2}_{j}$, which is unknown.   The   optimal tuning parameters of the CLIME and SCIO depend on $\|\bTheta\|_{1}$, which is unknown.

\subsubsection{The TIGER method}
To handle the challenge of tuning parameter selection,  \cite{LiuWang12} propose the {TIGER} (\underline{T}uning-\underline{I}nsensitive \underline{G}raph \underline{E}stimation and \underline{R}egression) method, which is asymptotically tuning-free and only requires very few efforts to choose the regularization parameter in finite sample settings.

 The  idea of TIGER  is to estimate  the precision matrix $\bTheta$ in a column-by-column fashion.  This idea has been adopted by many methods described in Section \ref{sec::columnregression}. These methods differ from each other mainly in how they solve the sparse regression subproblem. 
The only difference between TIGER and these methods is that TIGER solves its column-wise  sparse regression problem using the SQRT-Lasso \citep{Wang:12}.

 The SQRT-Lasso is a penalized optimization algorithm for solving high dimensional linear regression problems. For a linear regression problem $\widetilde \bY = \widetilde \Xb \bbeta+\bepsilon$, where $\widetilde \bY \in \mathbb{R}^T$ is the response vector, $\widetilde \Xb\in\mathbb{R}^{T\times p}$ is the design matrix,  $\bbeta \in \mathbb{R}^{p}$ is the vector of unknown coefficients, and
 $\bepsilon \in \mathbb{R}^T$ is the noise vector. The SQRT-Lasso estimates $\bbeta$ by solving
\begin{equation}
\hat{\bbeta}=\arg\min_{\bbeta\in\mathbb{R}^{p}}\Bigl\{ \frac{1}{\sqrt{T}}\|\widetilde \bY- \widetilde \Xb\bbeta\|_2+\lambda\|\bbeta\|_1\Bigr\},  \nonumber
\end{equation}
where $\lambda$ is a tuning parameter. It is shown in \cite{Wang:12}  that the choice of $\lambda$ for the SQRT-Lasso method is asymptotically universal in the sense that it does not depend on any unknown parameters such as the noise variance. In contrast, most of other methods, including the Lasso and Dantzig selector, rely heavily on variance of the noise. Moreover, the SQRT-Lasso method achieves near oracle performance for the estimation of  $\bbeta$.

 In \cite{LiuWang12}, they show that the objective function of the scaled-Lasso is a variational upper bound of the SQRT-Lasso. Thus the TIGER method is numerically equivalent to the method in  \cite{Zhang12}. However, the SQRT-Lasso is more amenable to theoretical analysis and allows us to simultaneously establish optimal rates of  convergence for the precision matrix estimation under many different norms.

In our setting, recall that $\Sbb$ is the sample covariance matrix of $\bY_t =(Y_{1t},\ldots, Y_{pt})'$. Let $\hat{\bGamma} = \mathrm{diag}(\Sbb)$ be a $p$-dimensional diagonal matrix with the diagonal elements be the same as those in $\Sbb$. 
Consider the marginally standardized variables
\begin{eqnarray}
\bZ := (Z_{1}, \ldots, Z_{p})'= \bY\hat{\bGamma}^{-1/2}.  \nonumber
\end{eqnarray}
By \eqref{gaussian.lin}, we have
\begin{equation}
\label{standgaussian.lin}
Z_j\hat{\bGamma}^{1/2}_{jj}=\balpha_j' \bZ_{\setminus j}\hat{\bGamma}^{1/2}_{\setminus j, \setminus j}+\epsilon_j.
\end{equation}
We define
\begin{eqnarray}
\bbeta_{j}:={\hat{\bGamma}^{1/2}_{\setminus j, \setminus j}}{\hat{\bGamma}^{-1/2}_{jj}} \balpha_{j}~~\text{and}~~\tau^{2}_{j} = \sigma^{2}_{j}\hat{\bGamma}^{-1}_{jj}.  \nonumber
\end{eqnarray}
Therefore, we have
\begin{eqnarray}
Z_{j} = \bbeta'_{j}\bZ_{\setminus j} + \hat{\bGamma}^{-1/2}_{jj}\epsilon_{j}. \label{eq::scaledModel}
\end{eqnarray}
We define $\hat{\Rb}$ to be the sample correlation matrix: $\hat{\Rb}:=\bigl(\diag(\Sbb)\bigr)^{-1/2}\Sbb\bigl(\diag(\Sbb)\bigr)^{-1/2}$.

Motivated by  the model in \eqref{eq::scaledModel}, we propose the following precision matrix estimator.
\begin{align}
\hline
\text{TIGER Estimator~~~~~~~~~~~~~~~~~~~~~~~~~~~~~~~}\nonumber\\
\text{For}~j=1,\ldots, p, ~\text{we estimate the $j^{\rm th}$ column of $\bTheta$ by solving}: \nonumber \\
\hat{\bbeta}_{j}:=\argmin_{\bbeta_{j}\in\mathbb{R}^{p-1}}\biggl\{\sqrt{1-2\bbeta'_{j}\hat{\Rb}_{\setminus j, j}  + \bbeta'_{j}\hat{\Rb}_{\setminus j, \setminus j} \bbeta_{j}} + \pi\sqrt{ \frac{\log p}{2T}}\bigl\|\bbeta_{j}\bigr\|_{1}\biggr\}, \label{eq::sqrtLasso}\\
\hat{\tau}_{j}:=\sqrt{1-2\hat{\bbeta}'_{j}\hat{\Rb}_{\setminus j, j}  + \hat{\bbeta}'_{j}\hat{\Rb}_{\setminus j, \setminus j} \hat{\bbeta}_{j}}, \label{eq::tauj}\\
\hat{\bTheta}_{jj} = \hat{\tau}^{-2}_{j}\hat{\bGamma}^{-1}_{jj}~~\text{and}~~\hat{\bTheta}_{\setminus j,j} = - \hat{\tau}^{-2}_{j}\hat{\bGamma}^{-1/2}_{jj}\hat{\bGamma}^{-1/2}_{\setminus j, \setminus j}\hat{\bbeta}_{j}. \nonumber\\
\hline \nonumber
\end{align}
Note that the first term in \eqref{eq::sqrtLasso} is just the square-root of the the sum of the square loss for the standardized variable under model \eqref{eq::scaledModel}; see \eqref{eq::SQRTLasso}.   We see that the TIGER procedure is  tuning free. If a symmetric precision matrix estimate is preferred, we conduct the following correction:
$\tilde{\bTheta}_{jk} = \min\bigl\{\hat{\bTheta}_{jk} , \hat{\bTheta}_{kj}  \bigr\}$ for all $k\neq j$.
Another symmetrization method is
\begin{eqnarray}
\tilde{\bTheta} = \frac{\hat{\bTheta} +\hat{\bTheta}'}{2}.  \nonumber
\end{eqnarray}
\cite{cai2011constrained} show that, if $\hat\bTheta$ is a good estimator, then $\tilde\bTheta$ will also be a good estimator: they achieve the same rates of convergence in the asymptotic settings.

Let $\Zb \in \mathbb{R}^{T\times p}$ be the normalized data matrix, i.e., $\Zb_{*j} = \Yb_{*j} \hat\bGamma^{-1/2}_{jj}$ for $j=1,\ldots, p$. An equivalent form of \eqref{eq::sqrtLasso} and \eqref{eq::tauj} is
\begin{align}
\hat{\bbeta}_{j} = \argmin_{\bbeta_{j}\in\mathbb{R}^{p-1}}\biggl\{\frac{1}{\sqrt{T}}\bigl\|\Zb_{*j} - \Zb_{*\setminus j} \bbeta_{j}\bigr\|_{2} + \lambda\bigl\|\bbeta_{j}\bigr\|_{1}\biggr\}, \label{eq::SQRTLasso}\\
\hat{\tau}_{j} =\frac{1}{\sqrt{T}}\bigl\|\Zb_{*j} - \Zb_{*\setminus j}\hat{\bbeta}_{j}\bigr\|_{2}. \label{eq::SQRTLasso2}
\end{align}
Once $\hat{\bTheta}$ is estimated, we can also estimate the graph $\hat{G}:=(V, \hat{E})$ based on the sparsity pattern of $\hat{\bTheta}_{jk}\neq 0$.

\cite{LiuWang12} establish the rates of convergence of the TIGER estimator $\hat{\bTheta}$ to  the true precision matrix $\bTheta$ under different norms.  Under the assumption that the condition number of $\bTheta$ is bounded by a constant, we have
\begin{eqnarray}
\bigl\|\hat{\bTheta} -\bTheta\bigr\|_{\max} = O_{P}\biggl(\|\bTheta\|_{1}\sqrt{\frac{\log p}{T}} \biggr). \label{eq::suprate}
\end{eqnarray}
Under mild conditions, the obtained rate in \eqref{eq::suprate} is minimax optimal  over the model class consisting of precision matrices with bounded condition numbers.

 The result in \eqref{eq::suprate} implies that  the Frobenious norm error  and spectral norm error between $\hat{\bTheta}$ and $\bTheta$ satisfy the following:
 let
 $s:=\sum_{j\neq k}1\left\{\bTheta_{jk}\neq 0\right\}$ be the number of nonzero off-diagonal elements of $\bTheta$; let $k:=\max_{i=1,\ldots, p}\sum_{j}1\{\bTheta_{ij}\neq 0\}$,
\begin{eqnarray}
\bigl\|\hat{\bTheta}-\bTheta\bigr\|_{\rm F} =  O_{P}\biggl(\|\bTheta\|_{1}\sqrt{\frac{(p+s) \log p}{T}} \biggr),\label{eq::Frobneius-rate}\\
\bigl\| \hat{\bTheta} - \bTheta\bigr\|_{2} =  O_{P}\left( k\|\bTheta\bigr\|_{2}\sqrt{\frac{\log p}{T}}\right). \label{spectra-rate}
\end{eqnarray}
The obtained rates in \eqref{spectra-rate} and \eqref{eq::Frobneius-rate} are  minimax optimal over the same model class as before.

\subsubsection{The EPIC method}

Another tuning-insensitive precision matrix estimation method is EPIC (\underline{E}stimating \underline{P}recision matr\underline{I}x with \underline{C}alibration),  proposed by \cite{zhao2014calibrated}. While TIGER can be viewed as a tuning-insensitive extension of the nodewise Lasso method proposed by \cite{Meinshausen06}, EPIC can be viewed as a tuning-insensitive extension of the CLIME estimator proposed by \cite{cai2011constrained}.  Unlike the TIGER method which relies on the normality assumption, the EPIC method can be used to handle both sub-Gaussian and heavy-tailed data. We postpone the details  of the EPIC method to Section 4 where we discuss robust estimators of covariance and precision matrices for heavy-tailed data.

\section{Robust  precision and covariance estimators}

The  methods introduced in Section 2 and Section 3 exploit the sample covariance matrix as input statistics. The theoretical justification of these methods  relies on the sub-Gaussian assumption of the data. However, many types of financial data are believed to follow the elliptical distributions, which are often heavy-tailed.  This section introduces a regularized rank-based framework for estimating  large precision and covariance matrices under  elliptical distributions.   First, we introduce a rank-based  precision matrix estimator which naturally handles heavy-tailness and  conducts parameter estimation under the elliptical models. Secondly, we introduce an adaptive rank-based covariance matrix estimator  which extends the generalized thresholding operator by adding an explicit eigenvalue constraint.   We also  provide interpretations of these rank-based estimators under the  more general elliptical copula model, which illustrates a tradeoff between  model flexibility and  interpretability.

Throughout this section, we assume  the data follow an elliptical distribution \citep{fang1990symmetric}, defined as below.

\begin{definition}[Elliptical Distribution]\label{stochastic-representation}
Given $\bmu \in \RR^p$ and a symmetric positive semidefinite matrix $\bSigma\in\RR^{p\times p}$ with $\mathrm{rank}(\bSigma) = r \leq p$, a $p$-dimensional random vector $\bY=(Y_1,...,Y_p)'$ follows an elliptical distribution with parameters $\bmu$, $\xi$, and $\bSigma$, denoted by $\bY\sim EC(\bmu,\xi,\bSigma)$, if $\bY$ has a stochastic representation
\begin{align}\label{stochastic-representation-equation}
\bY \mathop{=}^{d} \bmu + \xi\Ab\ub,
\end{align}
where $\xi \geq 0$ is a continuous random variable independent of $\ub$. Here $\ub \in \mathbb{S}^{r-1}$ is uniformly distributed on the unit sphere in $\RR^r$, and $\bSigma = \Ab\Ab'$.
\end{definition}
For notation convenience, we use $\xi$ instead the distribution of $\xi$ in the notation $EC(\bmu,\xi,\bSigma)$.  Note that the model in \eqref{stochastic-representation-equation}  is not identifiable since we can  rescale $\Ab$ and $\xi$  without changing the distribution. In this section, we require $\EE(\xi^2) <\infty$ and ${\rm rank}(\bSigma)=p$ to ensure  the existence of the inverse of $\bSigma$. In addition, we impose an identifiability condition $\EE(\xi^2) = p$ to ensure that $\bSigma$ is the covariance matrix of $\bY$. We still denote $\bTheta:=\bSigma^{-1}$.

\subsection{Robust precision matrix estimation}

To estimate  $\bTheta$,  our key observation is that  the covariance  matrix $\bSigma$ can be decomposed as  $\bSigma = \Db\Rb\Db$, where $\Rb$ is the Pearson's correlation matrix, and $\Db = {\rm diag}(\sigma_1,...,\sigma_p)$ where $\sigma_j$ is the standard deviation of $Y_j$. Since $\Db$ is diagonal, we can represent the precision matrix as $\bTheta = \Db^{-1}\bDelta\Db^{-1}$, where $\bDelta = \Rb^{-1}$ is the inverse correlation matrix. Based on this relationship,  the EPIC method of \cite{zhao2014calibrated} has three steps:   first obtain estimators $\hat{\Rb}$ and $\hat{\Db}$ for $\Rb$ and $\Db$; then apply a calibrated inverse correlation matrix estimation procedure on $\hat{\Rb}$ to obtain $\hat{\bDelta}$, an estimator for $\bDelta$. Finally,    assemble $\hat{\bDelta}$ and $\hat{\Db}$ to obtain a sparse precision matrix estimator $\hat{\bTheta}$.

 For light-tailed distributions (e.g., Gaussian or sub-Gaussian), we can directly use the sample correlation matrix and sample standard deviation to estimate the matrices $\Rb$ and $\Db$. However,  for  heavy-tailed elliptical data, the sample correlation matrix and standard deviation estimators are inappropriate.  Instead,   we exploit a combination of the transformed Kendall's tau estimator and Catoni's M-estimator, which will be explained in details in the following subsections.

\subsubsection{Robust estimation of correlation matrix}

To estimate $\Rb$, we adopt a transformed Kendall's tau estimator proposed in \cite{fang1990symmetric}. Define the population Kendall's tau correlation between $Y_{jt}$ and $Y_{kt}$ as
\begin{align*}
\tau_{kj} = \PP\left((Y_{jt}-\tilde{Y}_{jt})(Y_{kt}-\tilde{Y}_k)>0\right) -\PP\left((Y_{jt}-\tilde{Y}_j)(Y_{kt}-\tilde{Y}_k)<0\right),
\end{align*}
where $\tilde{Y}_j$ and $\tilde{Y}_k$ are independent copies of $Y_{jt}$ and $Y_{kt}$ respectively. For elliptical distributions, it is a well known result  that $\Rb_{kj}$ and $\tau_{kj}$ satisfy\footnote{More details can be found in \cite{fang1990symmetric}.}
\begin{align}\label{population-transformation}
\Rb = [\Rb_{kj}] = \left[\sin\left(\frac{\pi}{2}\tau_{kj}\right)\right].
\end{align}
The sample version Kendall's tau statistic between $Y_j$ and $Y_k$ is
\begin{align*}
\hat{\tau}_{kj} = \frac{2}{T(T-1)} \sum_{t<t'}\sign\Big((Y_{kt} - Y_{kt'})(Y_{jt} - Y_{jt'})\Big)
\end{align*}
for all $k \neq j$, and $\hat{\tau}_{kj}=1$ otherwise. We can   plug $\hat{\tau}_{kj}$ into  \eqref{population-transformation} and obtain a  rank-based correlation matrix estimator
\begin{align}\label{eq::Z}
\hat{\Rb} = [\hat{\Rb}_{kj}] = \left[\sin\left(\frac{\pi}{2}\hat{\tau}_{kj}\right)\right].
\end{align}

\subsubsection{Robust estimation of standard deviations}

To estimate $\Db$, we exploit an M-estimator proposed by \cite{catoni2012challenging}. Specifically, let
$\psi(t) = \sign(t)\cdot\log(1+|t|+t^2/2)$ be a univariate function where $\sign(0) = 0$. Let $\hat{\mu}_j$ and $\hat{m}_j$ be the estimators of $\EE Y_{jt}$ and $\EE Y_{jt}^2$ by solving the following two estimating equations:
\begin{align}
&\sum_{t=1}^T\psi\left((Y_{jt}-{\mu}_{j})\sqrt{\frac{2}{T K_{\max}}}\right) = 0,\label{catoni-equation1}\\
&\sum_{t=1}^T\psi\left((Y_{jt}^2-{m}_{j})\sqrt{\frac{2}{T K_{\max}}}\right) = 0,\label{catoni-equation3}
\end{align}
where $K_{\max}$ is an upper bound of $\max_j\Var(Y_{jt})$ and $\max_j\Var(Y_{jt}^2)$. We assume $K_{\max}$ is known. \cite{catoni2012challenging} shows that the solutions to \eqref{catoni-equation1} and \eqref{catoni-equation3} must exist and can be efficiently solved using the Newton-Raphson algorithm \citep{stoer1993introduction}. Once  $\hat{m}_j $ and $\hat{\mu}_j$ are obtained, we estimate the marginal standard deviation
$\sigma_j$ by
\begin{align}\label{catoni}
\hat{\sigma}_{j} = \sqrt{\max\bigl\{\hat{m}_j - \hat{\mu}_j^2, K_{\min}\bigr\}},
\end{align}
where $K_{\min}$ is a lower bound of $\min_j\sigma_j^2$ and is assumed to be known.

Compared to the sample covariance matrix,  a remarkable property of $\hat{\Rb}$ and $\hat{\sigma}_{j}$ is that they concentrate to their population quantities exponentially fast even for heavy-tailed data. More specifically, \cite{Liu12} show that
\begin{align}
\|\hat{\Rb} - \Rb\|_{\max} = O_P\Bigl(\sqrt{\frac{\log p}{T}} \Bigr)~~\text{and}~~\max_{1\leq j \leq p}\bigl|\hat{\sigma}_{j}  - \sigma_j \bigr| = O_P\Bigl(\sqrt{\frac{\log p}{T}} \Bigr).
\end{align}
In contrast, the sample correlation matrix and sample standard deviation do not have the above properties for heavy-tailed data.

\subsubsection{The EPIC method for inverse correlation matrix estimation}

Once $\hat{\Rb}$ and $\hat{\Db}$ are obtained, we need to estimate the inverse correlation matrix $\bDelta=\Rb^{-1}$. In this subsection, we introduce the EPIC method for estimating $\bDelta$, which estimates the $j^{\rm th}$ column of $\bDelta$ by plugging the transformed Kendall's tau estimator $\hat{\Rb}$ into the  convex program,
\begin{align}\label{EPIC}
(\hat{\bDelta}_{*j},\hat{\tau}_j)=\argmin_{\bDelta_{*j},\tau_j}~\norm{\bDelta_{*j}}_1+\frac{1}{2}\tau_j,~~
{\rm s.t.}~\norm{\hat{\Rb}\bDelta_{*j}-\Ib_{*j}}_{\infty} \leq \lambda\tau_j,~\norm{\bDelta_{*j}}_1\leq\tau_j.
\end{align}
Here $\tau_j$ serves as an auxiliary variable which ensures that we can use the same regularization parameter $\lambda$ for estimating different columns of $\bDelta$ \citep{gautier2011high}. Both the objective function and constraints in \eqref{EPIC} contain $\tau_j$, which ensures that $\tau_j$ is bounded.  \cite{zhao2014calibrated} show that  the regularization parameter $\lambda$ in \eqref{EPIC} does not depend on the unknown quantity $\bDelta$. Thus we can use the same $\lambda$ to estimate different columns of $\bDelta$.

The optimization problem in \eqref{EPIC} can be equivalently formulated as a linear program.  For notational simplicity, we omit the index $j$ in \eqref{EPIC}. We denote $\bDelta_{*j},~\Ib_{*j}$, and $\tau_j$ by $\bgamma,~\eb$, and $\tau$ respectively.  Let $\bgamma^+$ and $\bgamma^-$ be the positive and negative parts of $\bgamma$. By reparametrizing $\bgamma = \bgamma^+ - \bgamma^-$, we rewrite \eqref{EPIC} as the following linear program
\begin{align}\label{EPIC-LP}
(\hat{\bgamma}^+,\hat{\bgamma}^-,\hat{\tau}) = &\argmin_{\bgamma^+,\bgamma^-,\tau}~~\one'\bgamma^+ +\one'\bgamma^- + c\tau\\
&\hspace{-0.2in}{\rm s.t.}~~\left[
\begin{array}{rrr}
\hat{\Rb} 	&-\hat{\Rb} 	&-\blambda\\
-\hat{\Rb} 	&\hat{\Rb} 	&-\blambda\\
\one' 	&\one' 		&-1\\
\end{array}
\right]
\left[
\begin{array}{c}
\bgamma^{+}\\
\bgamma^{-}\\
\tau
\end{array}
\right]
\leq
\left[
\begin{array}{c}
\eb\\
-\eb\\
0
\end{array}
\right],\nonumber\\
&\hspace{0.2in}\bgamma^+ \geq \zero,~\bgamma^- \geq \zero,~\tau \geq 0,\nonumber
\end{align}
where $\blambda = \lambda\one$. The optimization problem in  \eqref{EPIC-LP} can be solved by  any linear program solver (e.g., the classical simplex method as suggested in \cite{cai2011constrained}). In particular, it can be efficiently solved using the parametric simplex method \citep{vanderbei2008linear}, which naturally exploits the underlying sparsity structure, and attains better empirical performance than a general-purpose solver.

\subsubsection{Symmetric precision matrix estimation}

Once we obtain  the inverse correlation matrix estimate $\hat{\bDelta}$, we can estimate $\bTheta$ by
\begin{align*}
\tilde{\bTheta} = \hat{\Db}^{-1}\hat{\bDelta}\hat{\Db}^{-1}.
\end{align*}
The EPIC method does not guarantee the symmetry of $\tilde{\bTheta}$. To obtain a symmetric estimator, we take an additional projection step:
\begin{align}
\hat{\bTheta} = \argmin_{\bTheta}\norm{\bTheta - \tilde{\bTheta}}_{*}~~~{\rm s.t.}~\bTheta=\bTheta',\label{symm-projection}
\end{align}
where $\norm{\cdot}_{*}$ can be the matrix $\ell_1$-, Frobenius, or elementwise max norm.
For both the Frobenius and elementwise max norms, \eqref{symm-projection} has a closed form solution
\begin{align*}
\hat{\bTheta} = \frac{1}{2}\left(\tilde{\bTheta}+\tilde{\bTheta}'\right).
\end{align*}
When using the matrix $\ell_1$-norm, the optimization problem in \eqref{symm-projection} does not have a closed-form solution. For this, we can exploit the smoothed proximal gradient algorithm to solve it. More details about this algorithm can be found in \cite{zhao2014calibrated}.

Consider a class of sparse symmetric matrices
\begin{align*}
&\cU(s,M,\kappa_u) = \Big\{\bDelta \in \RR^{p \times p} ~\Big|~\bDelta \succ 0,~\max_{j}\sum_{k}1\left\{\bDelta_{kj}\neq 0\right\} \leq s,\norm{\bDelta}_{1} \leq M,\Lambda_{\max}(\bDelta) \leq \kappa_u\Big\},
\end{align*}
where $\kappa_u$ is a constant, and $(s, p, M)$ may scale with the sample size $T$. Under some mild conditions, \cite{zhao2014calibrated} show that if we take $\lambda = \kappa_1\sqrt{(\log p)/T}$ and choose the matrix $\ell_1$-norm as $\norm{\cdot}_{*}$ in \eqref{symm-projection}, then for large enough $T$, we have
\begin{align}
\norm{\hat{\bTheta}-\bTheta}_2 = O_P\Bigl( M \cdot s\sqrt{\frac{\log p}{T}}\Bigr).\label{p-bound}
\end{align}
Moreover, if we choose the Forbenius norm as $\norm{\cdot}_{*}$ in \eqref{symm-projection}, then for large enough $T$,
\begin{align}
 \frac{1}{p}\norm{\hat{\bTheta}-\bTheta}^2_{\rm F} = O_P\Bigl(M^2\frac{s\log p}{T} \Bigr).\label{f-bound}
\end{align}

\subsection{Robust covariance matrix estimation}

In this subsection, we consider the problem of estimating the covariance matrix $\bSigma$ under the elliptical model \eqref{stochastic-representation-equation}. Similar to Section 2, we impose sparsity assumption on $\bSigma$.  To estimate $\bSigma$,  \cite{Liu2014EC2} introduce a regularized rank-based estimation method  named EC2 (\underline{E}stimation of \underline{C}ovariance  with  \underline{E}igenvalue \underline{C}onstraints), which can be viewed as an extension of the generalized thresholding operator \citep{rothman2009generalized}. The EC2 estimator can be formulated as the solution to  a convex program which ensures the  positive definiteness of the estimated covariance matrix. Unlike most existing methods, the EC2 estimator explicitly constrains the smallest eigenvalue of the estimated covariance matrix.

\subsubsection{The EC2 Estimator}

Recall that  $\bSigma = \Db\Rb\Db$. Similar to the EPIC method,  we calculate the EC2  estimator in three steps: In the first step, we  obtain robust estimators $\hat{\Rb}$ and $\hat{\Db}$ for $\Rb$ and $\Db$.  In the second step, we  apply an optimization  procedure on $\hat{\Rb}$ to obtain  $\hat{\Rb}^{\rm EC2}$, a sparse estimator for $\Rb$. In the third step,  we assemble $\hat{\Rb}^{\rm EC2}$ and $\hat{\Db}$ to obtain the final sparse covariance matrix estimator $\hat{\bSigma}=\hat{\Db}\hat{\Rb}^{\rm EC2}\hat{\Db}$. Specifically, we calculate $\hat{\Rb}$ and $\hat{\Db}$ as in \eqref{eq::Z} and \eqref{catoni}. In the following, we focus on explaining how to obtain $\hat{\Rb}^{\rm EC2}$ based on $\hat{\Rb}$.

Recall that $\hat{\Rb}$ is the transformed Kendall's tau  matrix, the $\hat{\Rb}^{\rm EC2}$ is calculated as
\begin{align}\label{EC2}
\hat{\Rb}^{\rm EC2} := \argmin_{\mbox{\scriptsize diag}(\Rb) =1}~\frac{1}{2}\|
\hat{\Rb}-\Rb\|_{\rm F}^2 + \lambda\norm{\Rb}_{1,\mathrm{off}}~~~\textrm{s.t.}~\tau \leq  \Lambda_{\min}(\Rb)
\end{align}
where $\lambda>0$ is a regularization parameter, and $\tau>0$ is a desired minimum eigenvalue lower bound of the estimator which is assumed to be known.  The EC2 method simultaneously conducts sparse estimation and guarantees the positive-definiteness of the solution. The equality constraint $\diag(\Rb) =1$ ensures that $\hat{\Rb}^{\rm EC2}$ is a correlation matrix. Once $\hat{\Rb}^{\rm EC2}$ is obtained, we convert it to the final covariance matrix estimator $\hat{\bSigma}$ as described above.  \cite{Liu2014EC2} prove the convexity of
 the formulation in \eqref{EC2}. Alternatively, one can apply thresholding on $\widehat \Rb$ to obtain a positive definite estimator.

\subsubsection{Asymptotic properties of the EC2 estimator}

To establish the asymptotic properties of the EC2 estimator, for $0\leq q < 1$, we consider the following class of sparse correlation matrices:
\begin{align}\label{bndeigen}
\cM\left(q, M_{p},\delta\right) :=\biggl\{\Rb:\max_{1\leq j\leq p} \sum_{k\neq j}\bigl|\Rb_{jk}\bigr|^{q} \leq M_{p}~\textrm{and}~\Rb_{jj}=1~\textrm{for~all}~ j,\Lambda_{\min}\bigl(\Rb\bigr)\geq\delta\biggr\}. \nonumber
\end{align}
We also define a class of covariance matrices:
\begin{align}
\cU(\kappa, q, M_{p}, \delta):=\biggl\{\bSigma:\max_{j}\bSigma_{jj}\leq \kappa~\textrm{and}~\Db^{-1}\bSigma\Db^{-1} \in \cM\left(q, M_{p},\delta\right)\biggr\},
\end{align}
where $\Db = {\rm diag}(\sqrt{\bSigma_{11}},...,\sqrt{\bSigma_{pp}})$. The definition of this class is similar to the  ``universal thresholding class'' defined by \cite{Bickel08a}.

Under the assumption  that the data follow an elliptical distribution, \cite{Liu2014EC2} show that, for large enough $T$, the EC2 estimator $\hat{\bSigma}$ satisfies
\begin{align} \label{eq::spectrarate}
\sup_{ \bSigma \in\cU(\kappa, q, M_{p}, \delta_{\min}) }\EE\bigl\|\hat{\bSigma}^{\mathrm{EC2}} - \bSigma\bigr\|_{2} \leq c_1\cdot M_{p} \Bigl(\frac{\log p}{T}\Bigr)^{\frac{1-q}{2}}.
\end{align}
\cite{Cai12} show that the rate in \eqref{eq::spectrarate} attains the minimax lower bound over the class $\cU(\kappa, q, M_{d}, \delta_{\min})$ under the Gaussian model.  Thus the EC2 estimator is asymptotically rate optimal under the flexible elliptical model with covariance matrix in $ \cU(\kappa, q, M_{d}, \delta_{\min})$.

\subsection{Extension to the elliptical copula family}

In Sections 4.1 and 4.2, we introduced the regularized rank-based covariance and precision matrix estimation methods by assuming the underlying distribution of $\bY=(Y_1,\ldots, Y_p)'$ is elliptical. In fact, these rank-based procedures also work within the more general transelliptical family \citep{liu2012transelliptical}, which is exactly the elliptical copula family but with different identifiability conditions.   More specifically, we  say $\bY=(Y_1,\ldots,Y_p)'$ follows a transelliptical distribution,
denoted by $\bY \sim TE(\bmu, \mathbf{\Sigma},\xi;f)$, if there exists
a set of strictly increasing functions $\{f_j\}_{j=1}^p$ such that
$f(\bm{Y})=(f_1(Y_1),\ldots,f_p(Y_p))'$ follows the elliptical
distribution $ EC(\bmu,\xi,\bSigma)$. To ensure the model is identifiable, \cite{liu2012transelliptical} impose the identifiability condition that, for $j\in\{1,\ldots, p\}$,
\begin{align}
\EE f_j(Y_j) = \EE Y_j~~\text{and}~~\mathrm{Var}(f_j(Y_j)) = \mathrm{Var}(Y_j).
\end{align}

As the Kendall's tau statistics in \eqref{eq::Z} are invariant under the monotonic transform,
the Kendall's tau statistics for the elliptical data $f(\bm{Y})$ are the same as those for the transellipitical data $\bm{Y}$.  Therefore, we do not need to estimate the monotonic transformations $f$ for computing the Kendall's tau.  On the other hand,
these monotonic transforms are not hard to estimate.  For example, for the Gaussian copula such that the marginal distribution $f_j(Y_j) \sim N(0, 1)$,  then based on the empirical distribution of the observed data $Y_j$ and the known marginal distribution $N(0,1)$, we can easily estimate $f_j$.

\begin{figure*}[t]
\includegraphics[width=0.95\textwidth,angle=0]{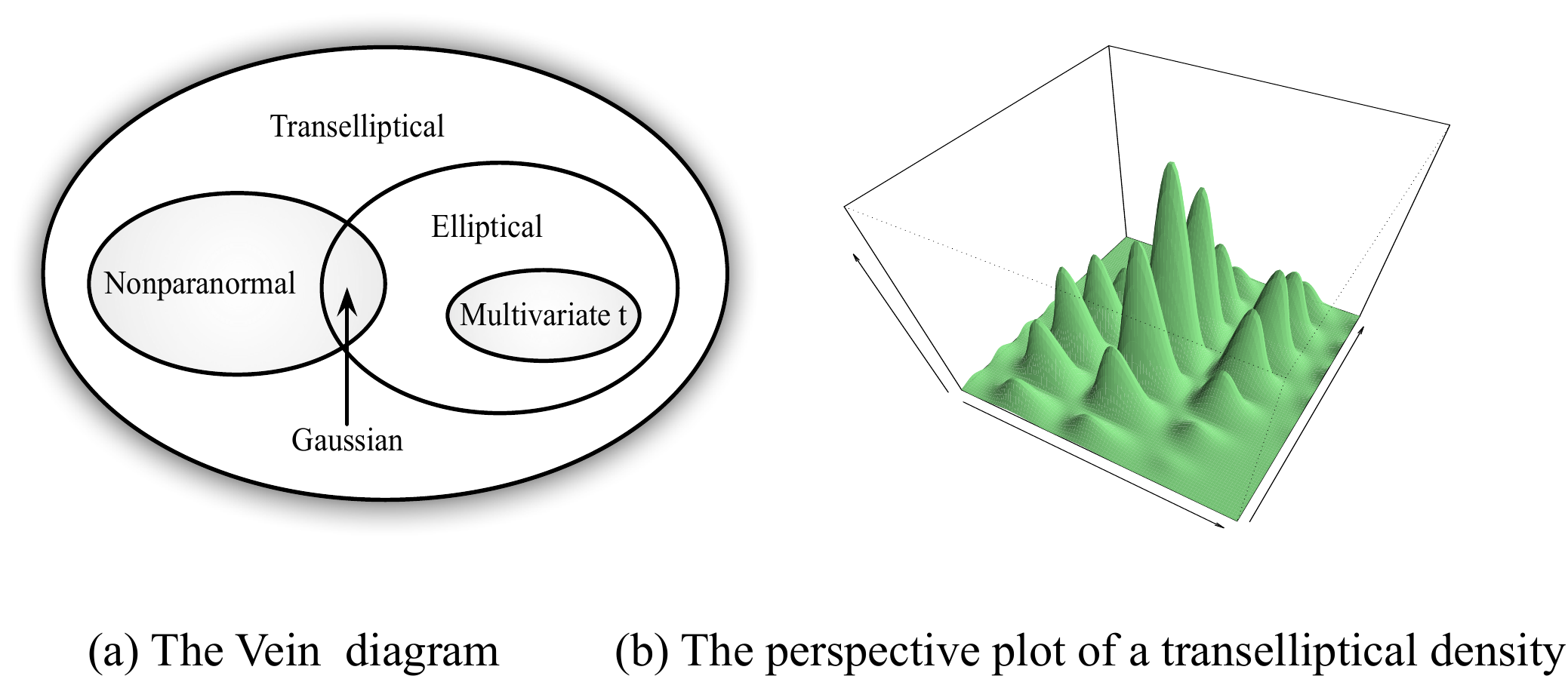}
 \caption{Transelliptical family. (a) The
   Vein diagram illustrating the relationships of the distribution
   families (The Nonparanomral family is equivalent to the Gaussian copula family).
   (b) The perspective plot of a transelliptical density.} \label{fig:vein}
\end{figure*}

\begin{figure*}[t]
\includegraphics[width=0.8\textwidth,angle=0]{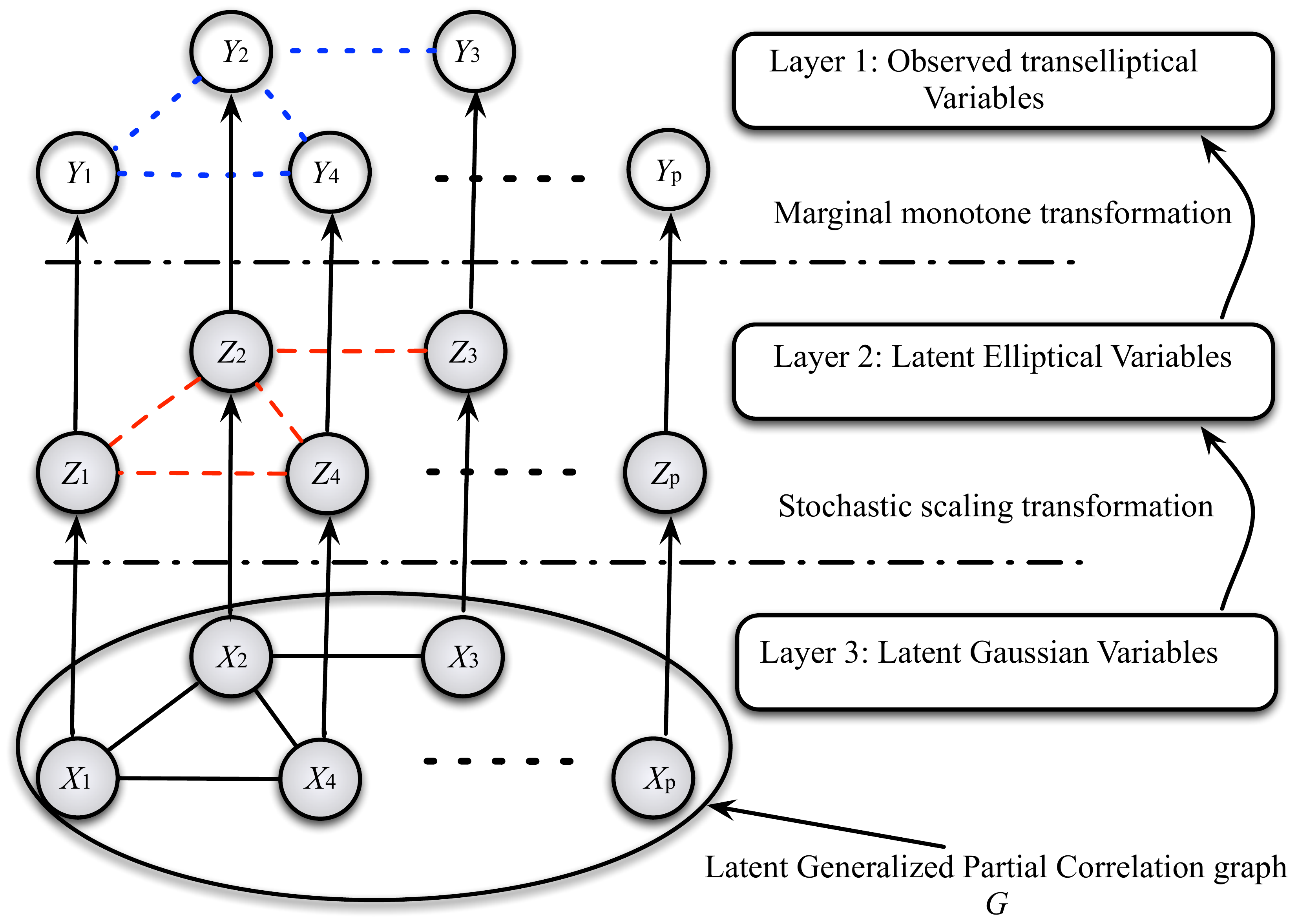}
 \caption{The hierarchical latent variable representation of the transelliptical graphical model
   with the latent variables grey-colored. Here the first layer is composed of observed $Y_j$'s,
   and the second and third layers are composed of latent variables $Z_j$'s and $X_j$'s. The solid
   undirected lines in the third layer encode the conditional independence graph of $X_1,\ldots,
   X_p$ (Adapted from a manuscript that is under review). } \label{fig:transelliptical}
\end{figure*}

Figure \ref{fig:vein}(a) illustrates the relationships
between the transelliptical, elliptical, and nonparanormal families \citep{Liu09a, Liu12}.  The nonparanormal family is a a proper subset of the transelliptical family.  We define
$\bY=(Y_1,\ldots,Y_p)'$ to be a nonparanormal distribution,
denoted by $\bY \sim NPN(\bmu, \mathbf{\Sigma},f)$, if there exists
a set of strictly increasing functions $\{f_j\}_{j=1}^p$ such that
$f(\bm{Y})=(f_1(Y_1),\ldots,f_p(Y_p))'$ follows the Gaussian
distribution $ N(\bmu,\bSigma)$. \cite{liu2012transelliptical} show that the intersection between the nonparanomral family and elliptical family is the Gaussian family.  Figure \ref{fig:vein}(b) visualizes the perspective plot of a bivariate transelliptical density with certain marginal transformations. The
transelliptical family is much richer than the elliptical family and its density function does not have to be symmetric.

The rank-based EPIC and EC2 methods can be directly applied to the transelliptical family. To
understand the semantics of a  transelliptical graphical model, \cite{liu2012transelliptical} proved that a
transelliptical distribution admits a three-layer hierarchical
latent variable representation as illustrated in
Figure \ref{fig:transelliptical}: The observed vector,
denoted by $\bm{Y}=(Y_1,\ldots,Y_p)'$ as presented in the first
layer, has a transelliptical distribution, and a latent random
vector, $\bm{Z}=(Z_1,\ldots,Z_p)'$ in the second-layer, is
elliptically distributed. Variables in the first and second layers
are related through the transformation $Z_j=f_j(Y_j)$ with $f_j$
being an unknown strictly increasing function. The latent vector $\bZ$ can be further
represented by a third-layer latent random vector
$\bm{X}=(X_1,\ldots,X_p)'$, which is a multivariate Gaussian
 with a covariance matrix $\mathbf{\Sigma}$ (called
{latent covariance matrix}) and an inverse covaraince matrix
$\mathbf{\Theta}=\mathbf{\Sigma}^{-1}$ (called {latent precision
matrix}).

We define the transelliptical graph $G=(V,E)$ with the
node set $V=\{1,\ldots,p\}$ and the edge set $E$ encoding the
nonzero entries of $\mathbf{\Theta}$. The  interpretations of the
graph $G$ are different for the variables in different layers: (i) For the observed
variables in the first layer, the absence of an edge between two
variables means the absence of a certain rank-based association (e.g., Kendall's tau) of
the pair given other variables; (ii) For the latent variables in
the second layer, the absence of an edge means the absence of the
conditional Pearson's correlation of the pair; (iii) For the third layer variables, the absence of an edge means the
conditional independence of the pair. Compared with the
Gaussian and elliptical graphical model, the
transelliptical graphical model has richer structure with more relaxed
modeling assumptions.  The three layers of hierarchy also reflects an interesting tradeoff between model flexibility and interpretability.  In the third layer, the model is the most restrictive Gaussian family, but we can get strong conditional independence arguments. In contrast, in the first layer, the model is the much  more flexible transelliptical family, but we can only get weaker conditional uncorrelatedenss (with respect to the rank correlation) statements.

Since the Kendall's tau statistic is monotone transformation invariant,   it is easy to see that the theory and methods  of the EPIC and EC2 procedures introduced in Section 4.1 and Section 4.2 are also applicable to the transelliptical distributions, though the interpretations of the fitted results are different (as explained in this section).

\section{ Factor model-based covariance estimation with observable factors}

Most of the aforementioned methods of estimating    $\bSigma$ assumes that  the covariance matrix is sparse.   Though this assumption is reasonable for many  applications, it is not always appropriate.  For example, financial stocks share the same market risks and hence their returns are highly correlated; all the genes from the same pathway may be co-regulated by a small amount of regulatory factors, which makes the gene expression data highly correlated; when genes are stimulated by cytokines, their expressions are also highly correlated. The sparsity assumption is obviously unrealistic in these situations.  

 In many applications, the responses of cross-sectional units often depend on a few common factors $\bbf$:
\begin{equation} \label{eq1.1add}
Y_{it}=\bbb_i'\bbf_t+u_{it}.
\end{equation}
Here  $\bbb_i$ is a  vector of factor loadings; $\bbf_t$  is a $K\times 1$ vector of common factors, and $u_{it}$  is the error term, usually called \textit{idiosyncratic  component}, uncorrelated with $\bbf_t$. Factor models have long been employed in financial studies, where $Y_{it}$ often represents the  excess returns of the $i$th asset (or stock) on time $t$. The literature includes, for instance, \cite{FF, chamberlain1983arbitrage, campbell1997econometrics}. It is also commonly used in macroeconomics for forecasting diffusion index (e.g., \cite{stock2002forecasting}). We allow $p,T\to\infty$ and that $p$ can grow much faster than $T$ does. In contrast, the number of factors $K$ needs to be either bounded or grows slowly.

This section introduces a method of estimating $\bSigma$ using factor models. We will focus on the case when the factors are observable.  The observable factor models are of considerable interest as they are often the case in empirical analyses in finance.


\subsection{Conditional sparsity}

The factor model (\ref{eq1.1add}) can be put in a matrix form as
\begin{equation}\label{e10}
    \bY_t=\Bb \bbf_t+{\bu_t}.
\end{equation}
where   $\Bb=({\bbb}_1,...,{\bbb}_p)'$ and $\bu_t=(u_{1t},...,u_{pt})'$.  We are interested in $\bSigma$,  the $p\times p$ covariance matrix of $\bY_t$, and its inverse $\bTheta=\bSigma^{-1}$, which are assumed to be time-invariant.   Under model (\ref{e10}) and the independence assumption between $\bbf_t$ and $\bu_t$,  $\bSigma$ is given by
\begin{equation}\label{e11}
\bSigma=\Bb\Cov(\bbf_t)\Bb'+\bSigma_u,
\end{equation}
where $\bSigma_u=(\sigma_{u,ij})_{p\times p}$ is the covariance matrix of $\bu_t$. Estimating  the covariance matrix $\bSigma_u$ of the idiosyncratic components $\{\bu_t\}$ is also important for  statistical inferences. 

\cite{fan2008high}  studied  model (\ref{e11}) when $p\to\infty$ possibly faster than $T$. They assumed $\bSigma_u$  to be a diagonal matrix, which corresponds to the classical ``strict factor model", and might be restrictive in practical applications.  On the other hand,  factor models are often only justified as being ``approximate", in which the  $Y_{1t},...,Y_{pt}$ are still mutually correlated given the factors, though the mutual correlations are weak.  This gives rise to the \textit{approximate factor model} studied by \cite{chamberlain1983arbitrage}.  In the approximate factor model, $\bSigma_u$  is a non-diagonal covariance matrix, and admits many small off-diagonal entries.

 In the decomposition (\ref{e11}), we assume $\bSigma_u$ to be sparse. This can be interpreted as the conditional sparse covariance model: Given the common factors $\bbf_1,...,\bbf_T$, the conditional (after taking out the linear projection on to the space spanned by the factors) covariance matrix of $\bY_t$ is sparse.   Let
  \begin{equation}
m_{u,p}=\begin{cases}\label{e1add}
\max_{i\leq p}\sum_{j=1}^p1\{\sigma_{u,ij}\neq0\},&\text{ if } q=0\\
\max_{i\leq p}\sum_{j=1}^p|\sigma_{u,ij}|^{q},&\text{ if } 0<q<1
\end{cases}.
\end{equation}
We require $m_{u,p}$ be either bounded or grow slowly as $p\to\infty$.   The conditional sparsity assumption is slightly stronger than those  of   the approximate factor model in \cite{chamberlain1983arbitrage}, but is still  a  natural assumption: the idiosyncratic components are mostly uncorrelated.   In contrast, note that    in the presence of common factors, $\bSigma$ itself is hardly a sparse matrix.

\subsection{Estimation}

When the factors are observable, one can estimate $\Bb$ by the ordinary  least squares (OLS): $\widehat\Bb=({\widehat\bbb}_1,...,{\widehat\bbb}_N)'$, where,
$$
\widehat\bbb_i=\arg\min_{{\bbb}_i}\frac{1}{T}\sum_{t=1}^T( Y_{it}-{\bbb}_i'{\bbf_t})^2,\quad i=1,...,N.
$$
Then, $\widehat{\bu}_{t} = \bY_t - \widehat \Bb \bbf_t$ is the residual vector at time $t$.
We then construct the residual covariance matrix as:
$$
\Sbb_u=\frac{1}{T}\sum_{t=1}^T\widehat\bu_t \widehat\bu_t'=(s_{u,ij}).
$$

Since $\bSigma_u$ is sparse, we now apply thresholding on $\Sbb_u$ to regularize the estimator.  Define
$$
\widehat\bSigma_u=(\widehat\sigma_{u,ij})_{p\times p},\hspace{1em}\widehat\sigma_{u, ij}=\begin{cases}
s_{u,ii}, & i=j;\\
h(s_{u,ij}; \omega_{T,ij}), & i\neq j.
\end{cases}
$$
Here $h(.;\omega_{T,ij})$ is a general thresholding rule as described in Section 2.  Both the adaptive thresholding and entry dependent thresholding can also be incorporated, by respectively setting $\omega_{T,ij}=\text{SE}(s_{u,ij})\omega_T$ and $\omega_{T,ij}= \sqrt{s_{u,ii}s_{u,jj}}\omega_T$, with
$$
\omega_T=CK\sqrt{\frac{\log p}{T}}
$$
for some $C>0$. As in the discussions in Section 2, $C>0$ can be chosen via cross-validation in a proper  range to guarantee the finite sample positive definiteness.


 The covariance matrix $\Cov({\bbf_t})$ can be estimated by the sample covariance matrix
 $$\widehat\Cov({\bbf_t})=\frac{1}{T}\sum_{t=1}^T(\bbf_t-\bar \bbf)(\bbf_t-\bar \bbf)', \quad \bar\bbf=\frac{1}{T}\sum_{t=1}^T\bbf_t,
$$
which does not require regularization since the number of factor is assumed to be small.  Therefore  we obtain a  substitution estimator:
$$
\widehat\bSigma=\widehat\Bb\widehat\Cov({\bbf_t})\widehat\Bb'+\widehat\bSigma_u.
$$
By the Sherman-Morrison-Woodbury formula, we estimate the precision matrix as
$$
\widehat\bSigma^{-1}=\widehat\bSigma_u^{-1}-\widehat\bSigma_u^{-1}\widehat\Bb[\widehat\Cov({\bbf_t})^{-1}+\widehat\Bb'\widehat\bSigma_u^{-1}\widehat\Bb]^{-1}\widehat\Bb'
\widehat\bSigma_u^{-1}.
$$

Under regularity conditions, \cite{FLM11} showed that when $m_{u, p}\omega_T^{1-q}\to0$,
 $$
 \|\widehat\bSigma_u-\bSigma_u\|_2=O_P(m_{u, p}\omega_T^{1-q}),\quad  \|\widehat\bSigma_u^{-1}-\bSigma_u^{-1}\|_2=O_P(m_{u, p}\omega_T^{1-q}),
 $$
On the other hand, it is difficult to obtain a satisfactory convergence rate for $\widehat\bSigma$ under either the operator or the Frobenius norm.  We illustrate this problem in the following example. Let $\zero_d$ be a $d$-dimensional row vector of zeros.

\begin{example}  Consider the specific case $K = 1$ with the known loading $\bB = \one_p$ and  $\bSigma_u=\Ib$. Then  $\bSigma = \Var(f_1) \one_p \one_p^T + \Ib$, where $\one_p$ denotes the $p$-dimensional column vector of ones with $\|\one_p\one_p'\|_2=p$, and we only need to estimate $\Var(f_1)$ using the sample variance.  Then, it follows that
$$
\|\widehat\bSigma-\bSigma\|_2=|\frac{1}{T}\sum_{t=1}^T(f_{1t}-\bar{f}_1)^2-\Var(f_{1t})|\cdot\|\one_p\one_p'\|_2,
$$
Therefore,  it follows from the central limit theorem that $\frac{\sqrt{T}}{p}\|\widehat\bSigma-\bSigma\|_2$ is asymptotically normal. Hence $\|\widehat\bSigma-\bSigma\|_2$ diverges  if $p\gg \sqrt{T}$, even for such a simplified toy model.

\end{example}

In the above toy example,  the bad rate of convergence is mainly due to the large quantity $\|\one_p\one_p'\|_2$, which comes from the high-dimensional factor  loadings.  In general,  the high-dimensional loading matrix accumulates many estimation errors.

On the other hand,  \cite{FLM11}  showed that we can obtain a good convergence rate when estimating $\bSigma^{-1}$:
$$\|\widehat\bSigma^{-1}-\bSigma^{-1}\|_2=O_P(m_{u, p}\omega_T^{1-q}).$$
Intuitively, the good performance of $\widehat\bSigma^{-1}$ follows from the fact that the eigenvalues of $\bSigma^{-1}$ are uniformly bounded, whereas  the leading  eigenvalues of $\bSigma$ may diverge fast.

\section{Factor models-based covariance estimation with latent factors}

In many empirical studies using factor models, the common factors are often latent, that is, they are unobservable.  In this case, the covariance matrix of $\bY_t$ has the same decomposition as before:
\begin{equation}\label{e13}
\bSigma=\Bb\Cov(\bbf_t)\Bb'+\bSigma_u,
\end{equation}
but  the latent factors also need to be estimated. Similar to the case of observable factors, the model can be assumed to be conditionally sparse, where $\bSigma_u$ is a sparse matrix but not necessarily diagonal. In this section we shall assume the number of factors to be bounded.

\subsection{The pervasive condition}

Note that unlike the classical factor analysis (e.g., \cite{lawley1971factor}), when $\bSigma_u$ is non-diagonal, the decomposition (\ref{e13}) is not identifiable under fixed $(p,T)$, since $\bY_t$ is the only observed data in the model. Here the identification means the separation of the low-rank part $\Bb\Cov(\bbf_t)\Bb'$ from $\bSigma_u$ in the decomposition (\ref{e13}). Interestingly,  however, the identification of $\Bb\Cov(\bbf_t)\Bb'$   can be achieved asymptotically, by letting $p\to\infty$ and requiring the eigenvalues of $\bSigma_u$ to be either uniformly bounded or grow slowly relative to $p$.

 What makes the ``asymptotic identification" possible is the following \textit{pervasive} assumption, which is one of the key conditions assumed in the literature (e.g., \cite{stock2002forecasting, bai2003inferential}):
 \begin{assumption}
The eigenvalues of the $K\times K$ matrix $p^{-1}\Bb'\Bb=\frac{1}{p}\sum_{i=1}^p\bbb_i\bbb_i'$ are uniformly bounded away from both zero and infinity, as $p\to\infty$.
 \end{assumption}

 When this assumption is satisfied, the factors are said to be ``pervasive".
 It requires the factors impact on most of the cross-sectional individuals. It then follows that the first $K$ eigenvalues of $\Bb\Cov(\bbf_t)\Bb'$ are bounded from below by $c\lambda_{\min}(\Cov(\bbf_t))p$ for some $c>0$,  and should grow fast with $p$. On the other hand,
\begin{equation}\label{e14}
 \|\bSigma_u\|_2\leq \max_{i\leq p}\sum_{j=1}^p|\sigma_{u,ij}|^{q}|\sigma_{u,ii}\sigma_{u,jj}|^{(1-q)/2}\leq m_{u,p}\max_{i\leq p}\sigma_{u,ii}^{1-q}.
 \end{equation}
 Hence  when $m_{u,p}$ grows slower than $O(p)$, the leading eigenvalues of the two components on the right hand side of (\ref{e13}) are well separated as $p\to\infty$. This guarantees that the covariance decomposition is asymptotically identified.
  Intuitively, as the dimension increases, the information about the common factors accumulates, while the information about the idiosyncratic components does not. This eventually distinguishes the factor components $\Bb\bbf_t$ from $\bu_t$.

  Below we shall introduce a principal component analysis (PCA) based method to estimate the covariance matrix.

\subsection{Principal Component and Factor Analysis}

Before introducing the estimator of $\bSigma$ in the case of latent factors, we first elucidate why PCA can be used for the factor analysis   when the number of variables is large.  First of all, note that even if $\Bb\Cov(\bbf_t)\Bb'$ is asymptotically identifiable, $\Bb$ and $\bbf_t$ are not separately identifiable, since the pair $(\Bb,\bbf_t)$ is  equivalent to the pair $(\Bb\Hb^{-1}, \Hb\bbf_t)$ for any $K\times K$ nonsingular matrix $\Hb$.  To resolve the ambiguity between $\Bb$ and $\bbf_t$, we impose the identifiability constraint that $\Cov(\bbf_t)=\Ib_K$ and that the  columns of $\Bb$ are orthogonal. Under this canonical form, it then follows from (\ref{e13}) that
$$
\bSigma=\Bb\Bb'+\bSigma_u.
$$

 Let $\widetilde\bbb_1,...,\widetilde\bbb_K$ be the columns of $\Bb$.  Since the columns of $\Bb$ are orthogonal,  $$
\Bb\Bb'\widetilde\bbb_j=\widetilde\bbb_j\|\widetilde\bbb_j\|^2_2,\quad \mbox{for } j\leq K.
$$
 Therefore, $\widetilde\bbb_1/\|\widetilde\bbb_1\|_2, \cdots, \widetilde\bbb_K/\|\widetilde\bbb_K\|_2$ are the eigenvectors of $\Bb\Bb'$, corresponding to the   largest $K$ eigenvalues   $\{\|\widetilde\bbb_j\|^2_2\}_{j=1}^K$; the rest $p-K$ eigenvalues  of $\Bb\Bb'$ are zeros. To guarantee the uniqueness (up to a sign change) of the leading eigenvectors, we also assume $\{\|\widetilde\bbb_j\|_2\}_{j=1}^K$ are distinct and sorted in a decreasing order.
 To see how large these eigenvalues are, note that the first $K$ eigenvalues of $\Bb\Bb'$ are the same as those of $\Bb'\Bb$. Hence it follows from the pervasive assumption  (Assumption 1)  that
 \begin{equation}\label{e15}
 \|\widetilde \bbb_j\|_2^2\geq cp,\quad j=1,...,K.
 \end{equation}

 Next, let us associate the leading eigenvalues of $\Bb\Bb'$ with those of $\bSigma$. Let $\lambda_1,...,\lambda_K$ denote the   $K$ largest  eigenvalues of $\bSigma$, and let  $\bxi_1,...,\bxi_K$ be the corresponding eigenvectors.
  Applying  Wely's theorem and the $\sin(\theta)$-theorem of \cite{davis1963rotation},   \cite{POET} showed
 $$
\|\bxi_j- \widetilde\bbb_j/\|\widetilde\bbb_j\|_2\|_2=O(p^{-1}\|\bSigma_u\|_2),\quad \text{ for all } j\leq K.
$$
 and
 $$
 |\lambda_j-\|\widetilde\bbb_j\|_2^2|\leq\|\bSigma_u\|_2, \text{ for } j\leq K\quad, |\lambda_j|\leq \|\bSigma_u\|_2,\text{ for } j>K.
 $$
 These results demonstrate:
 \begin{enumerate}
 \item The leading eigenvectors of $\bSigma$ are approximately equal to the normalized columns of $\Bb$, as $p\to\infty$.  In other words, the factor analysis and the principal analysis are approximately the same.
 \item  The leading eigenvalues of $\bSigma$ grow at rate $O(p)$. This can be seen from applying the triangular inequality and (\ref{e14}), (\ref{e15}):
 $$
 \lambda_j>\|\widetilde\bbb_j\|_2^2- |\lambda_j-\|\widetilde\bbb_j\|^2|\geq cp-m_{u,p}\max_{i\leq p}\sigma_{u,ii}^{1-q},\quad \forall j=1,...,K.
 $$
 \item  The latent factor $f_{jt}$ is  approximately $\bxi_j'\bY_t/\sqrt{\lambda_j}$ for $j=1,...,K$.  To see this, left-multiplying $\widetilde\bbb_j'/\|\widetilde\bbb_j\|_2^2$ to $\bY_t=\Bb\bbf_t+\bu_t$, and noting that the columns of $\Bb$ are orthogonal, we have
 $$
 f_{jt}=\widetilde\bbb_j'\bY_t/\|\widetilde\bbb_j\|_2^2-\widetilde\bbb_j'\bu_t/\|\widetilde\bbb_j\|_2^2.
 $$
 The second term on the right is the weighted average of noise $\bu_t$ over all $p$ individuals and hence typically negligible when $p$ is large.  The first term is
 $$
\frac{ \widetilde\bbb_j'\bY_t}{\|\widetilde\bbb_j\|_2^2}=\frac{ \widetilde\bbb_j'/\|\widetilde\bbb_j\|_2\bY_t}{\|\widetilde\bbb_j\|_2} \approx \frac{ \bxi_j'\bY_t}{ \sqrt{\lambda_j}}.
 $$
 Hence as $p\to\infty$, $f_{jt}\approx  \bxi_j'\bY_t/\sqrt{\lambda_j}$.
 \end{enumerate}

 Therefore, we conclude that the first $K $ eigenvalues of $\bSigma$ are very spiked, whereas the remaining eigenvalues are either bounded or grow slowly. In addition, both the latent factors and loadings can be approximated using the eigenvalues and eigenvectors of $\bSigma$ and $\bY_t$. This builds the connection between the PCA and  high-dimensional  factor models.

 \subsection{POET estimator}

 \cite{POET} proposed a nonparametric estimator of $\bSigma$ when the factors are unobservable, named POET (Principal Orthogonal complEment Thresholding). To motivate their estimator, note that $\Bb\Bb'=\sum_{j=1}^K\widetilde\bbb_j\widetilde\bbb_j'$. From the discussions of the previous subsection, heuristically we have
 $$
\sum_{j=1}^K\widetilde\bbb_j\widetilde\bbb_j'\approx \sum_{j=1}^K\lambda_j\bxi_j\bxi_j'.
 $$
 In fact, it can be formally proved that
 $$
 \|\Bb\Bb'-\sum_{j=1}^K\lambda_j\bxi_j\bxi_j'\|_{\max}=O(p^{-1/2}),
 $$
 which can be understood as  the (asymptotic) identification  for $\Bb\Bb'$.
 In addition, note that $\bSigma$ has  the spectral decomposition $\bSigma=\sum_{j=1}^p\lambda_j\bxi_j\bxi_j'$ and the factor decomposition $\bSigma=\Bb\Bb'+\bSigma_u$. Therefore,
 $$
 \bSigma_u\approx  \sum_{j=K+1}^p\lambda_j\bxi_j\bxi_j'.
 $$
 Under the conditional sparsity assumption, $\sum_{j=K+1}^p\lambda_j\bxi_j\bxi_j'$ is approximately a sparse matrix.  One can then estimate $\bSigma_u$ by thresholding the sample analogue of $\sum_{j=K+1}^p\lambda_j\bxi_j\bxi_j'$.

Specifically, the POET estimator is defined as follows.   Let $\widehat\lambda_1\geq\widehat\lambda_2\geq\cdots\geq\widehat\lambda_p$ be the ordered eigenvalues of the sample covariance matrix $\Sbb$, and $\widehat\bxi_1,...,\widehat\bxi_p$ be the corresponding eigenvectors.  Then the sample covariance has the following spectral decomposition:
$$
\Sbb=\sum_{i=1}^K\widehat\lambda_i\widehat\bxi_i \widehat\bxi_i'+\Sbb_u,
$$
where $\Sbb_u=\sum_{k=K+1}^p\widehat\lambda_k\widehat\bxi_k \widehat\bxi_k'=(s_{u,ij})$,  called ``the principal orthogonal complement".   We apply the generalized thresholding rule on $\Sbb_u$. Define $$
\widehat\bSigma_u=(\widehat\sigma_{u,ij})_{p\times p},\hspace{1em}\widehat\sigma_{u, ij}=\begin{cases}
s_{u,ii}, & i=j;\\
h(s_{u,ij}; \widetilde\omega_{T,ij}), & i\neq j.
\end{cases}
$$
For instance, the  entry dependent thresholding sets   $\widetilde\omega_{T,ij}= \sqrt{s_{u,ii}s_{u,jj}}\widetilde\omega_T$. Importantly, $\widetilde\omega_T$ is different from before when the factors are latent, and should be set to
$$
\widetilde\omega_T=C\left(\sqrt{\frac{\log p}{T}}+\frac{1}{\sqrt{p}}\right).
$$
It was then shown by \cite{POET} that
$$
\max_{i,j\leq p}|s_{u,ij}-\sigma_{u,ij}|=O_P(\widetilde\omega_T).
$$
 The extra term $\frac{1}{\sqrt{p}}$ in $\widetilde\omega_T$ is the price paid for not knowing the latent factors, and is negligible when $p$ grows faster than $T$. Intuitively, when the dimension is sufficiently large, the latent factors can be estimated accurately enough as if they were observable.

 The POET estimator of $\bSigma$ is then defined as:
\begin{equation}\label{eq2.4}
\widehat\bSigma_K=\sum_{i=1}^K\widehat\lambda_i\widehat\bxi_i\widehat\bxi_i'+\widehat\bSigma_u.
\end{equation}
This estimator is optimization-free and is very easy to compute.

Note that $\widehat\bSigma_K$ requires the knowledge of $K$, which is the number of factors and practically unknown. There has been a large literature on determining the number of factors and many consistent estimators have been proposed,  such as  \cite{BN02, alessi2010improved, hallin2007determining},  and \cite{ahn2013eigenvalue}. In addition, numerical studies in \cite{POET} showed that the covariance estimator is robust to over-estimating $K.$ Therefore, in practice, we can also choose a relatively large number for $K$ even if it is not a consistent estimator of the true number of factors. In the sequel, we suppress the subscript $K$, and simply write   $\widehat\bSigma$ as the POET estimator.

\subsection{Asymptotic Results}

Under the conditional sparsity assumption and some regularity conditions,  \cite{POET} showed that when  $\widetilde\omega_T^{1-q}m_{u,p}\to0$, we have
$$
    \|\widehat\bSigma_u-\bSigma_u\|_2=O_P\left( \widetilde\omega_T^{1-q}m_{u,p} \right), \quad \|\widehat\bSigma_u^{-1}-\bSigma_u^{-1}\|_2=O_P\left(\widetilde\omega_T^{1-q}m_{u,p}\right).$$

On the other hand, the problem of bad rate of convergence for $\bSigma$ is still present, because the first $K$ eigenvalues of $\bSigma$ grow with $p$. We can further illustrate this point in the following example (taken from \cite{POET}):

\begin{example}   \label{exam31}
Consider an ideal case  where we know the spectrum except for the first eigenvector  of $\bSigma$, and assume that  the largest eigenvalue  $\lambda_1\geq cp$ for some $c>0$.  Let $\widehat\bxi_1$ be the estimated first eigenvector and define the covariance estimator
$
\widehat\bSigma=\lambda_1\widehat\bxi_1\widehat\bxi_1'+\sum_{j=2}^{p}\lambda_j\bxi_j\bxi_j'.
$
Assume that $\widehat\bxi_1$ is a good estimator in the sense that $\|\widehat\bxi_1-\bxi_1\|^2=O_p(T^{-1})$. However,
$$
\|\widehat\bSigma-\bSigma\|_2=\|\lambda_1(\widehat\bxi_1\widehat\bxi_1'-\bxi_1\bxi_1')\|_2=\lambda_1O_p(\|\widehat\bxi-\bxi\|_2)=O_p(\lambda_1T^{-1/2}),
$$
which can diverge when $T=O(p^2)$.
\end{example}

Similar to the case of observable factors, we  can   estimate the precision matrix with a satisfactory rate under the operator norm.    The intuition still follows from the fact that $\bSigma^{-1}$ has bounded eigenvalues. Indeed, \cite{POET} showed that $\widehat\bSigma^{-1}$ has the same rate of convergence as that of $\widehat\bSigma_u^{-1}$.

\section{Structured factor models}

\subsection{Motivations}\label{s31}
In the usual  asymptotic analysis for factor models, accurate estimations of the space spanned by the eigenvectors of $\bSigma$ require a relatively large $T$. In particular, the individual loadings can be estimated no faster than $O_P(T^{-1/2})$. But data sets of large sample size are not always available. Often we face    the ``high-dimensional-low-sample-size" (HDLSS) scenario, as described  in \cite{jung2009pca}.  This is particularly the case in financial studies of asset returns, as their dynamics can vary substantially over a longer time horizon.  Therefore, to capture the current market condition, financial analysts wish to use short time horizon to infer as good as possible the risk factors as well as their associated loading matrix.  To achieve this, we need additional data covariate information and modeling of the factor loadings.

Suppose that there is a $d$-dimensional vector of  observed covariates associated with the $i^{th}$ variable: $\bX_i=(X_{i1} , \cdots,  X_{id})$, which is independent of $u_{it}$.  For instance,  in financial applications, $\bX_i$ can be a vector of firm-specific characteristics (market capitalization, price-earning ratio, etc);
in health studies, $\bX_i$ can be individual characteristics (e.g. age, weight, clinical and genetic information).
To incorporate the information carried by the observed characteristics,   \cite{CL07} and \cite{CMO} model explicitly the loading matrix as a function of covariates $\bX$.  This reduces significantly the number of paramters in $\bB$.  Specifically, they proposed and studied  the following semi-parametric factor model:
\begin{equation} \label{eq43}
Y_{it}=\sum_{k=1}^K g_k(\bX_{i})f_{kt}+u_{it}, \quad i=1,\cdots,p, t=1,\cdots,T.
\end{equation}
Here  $g_k (\bX_i)$ is an unknown function of the characteristics and they assume further the additive modeling
\begin{equation} \label{eq44}
    g_k(\bX_i) = g_{k1}(X_{i1}) + \cdots + g_{kd} (X_{id}).
\end{equation}

\cite{fan2014projected} recognized that the above semi-parametric model \eqref{eq43} might  be restrictive for applications, as we do not expect that the covariates capture completely the factor loadings.  They extend the model to the following more flexible semiparametric mixed effect model:
\begin{equation}\label{eq4}
Y_{it}=\sum_{k=1}^K [g_k(\bX_{i})+\gamma_{ik}]f_{kt}+u_{it}, \quad i=1,\cdots,p, t=1,\cdots,T.
\end{equation}
Here  $\gamma_{ik}$ is an unobservable random component with mean zero.  They developed econometric techniques to test the model specifications (\ref{eq43}) and (\ref{eq4}).  Their empirical results, using the returns of the components of the S\&P500 index and 4 exogenous variables (size, value, momentum, and volatility) as in \cite{CMO}, provide stark evidence that model (\ref{eq43}) can not be validated empirically whereas (\ref{eq4}) is consistent with the empirical data.

\subsection{Projected PCA}

The basic idea of projected PCA is to smooth the observations $\{Y_{it}\}_{i=1}^p$ for each given day $t$ against its associated covariates $\{\bX_{i}\}_{i=1}^p$.  More specifically, let $\{\hat{Y}_{it}\}_{i=1}^p$ be the fitted value after run a regression of $\{Y_{it}\}_{i=1}^p$ against $\{\bX_{i}\}_{i=1}^p$ for each given $t$.  The regression model can be the usual linear regression or additive regression model \eqref{eq44}.  This results in a smooth or projected observation matrix $\hat{\Yb}$, which will also be denoted by $\Pb \Yb$.  The projected PCA is then to run PCA based on the projected data $\hat{\Yb}$.

To provide the rationale behind this idea, we now generalize  model \eqref{eq4} further to illustrate the idea behind the projected PCA.  Specifically, consider the factor model
$$
\Yb=\Bb\Fb'+\Ub
$$
where $\Yb$ and $\Ub$ are $p\times T$ matrices of $y_{it}$ and $u_{it}$.  Suppose that there is a $d$-dimensional vector of  observed covariates associated with the $i^{th}$ variable: $\bX_i=(X_{i1} , \cdots,  X_{id})$, which is independent of $u_{it}$.
For  a pre-determined $J$, let $\phi_1,...,\phi_J$ be a set of basis functions.  Let $\phi(\bX_i)'=(\phi_1(X_{i1}),....,\phi_J(X_{i1}),....,\phi_J(X_{id}))$ and $\Phi(\Xb)=(\phi(\bX_1),...,\phi(\bX_p))'$ be a $p\times (Jd)$ matrix of the sieve-transformed $\Xb$. Then the projection matrix on the space spanned by $\Xb=(\bX_1,...,\bX_p)$ can be  taken as
$$
\Pb=\Phi(\Xb)(\Phi(\Xb)'\Phi(\Xb))^{-1}\Phi(\Xb)'.
$$
This corresponds to modeling $g_k(\bX_i)$ in \eqref{eq4} by the additive model \eqref{eq44}
and approximating each term using the series expansion.
 The projected data $\Pb \Yb$ is the fitted value of the additive model \eqref{eq44} with basis functions $\phi_1,...,\phi_J$:
$$
    Y_{it} = \sum_{k=1}^K [ \sum_{j=1}^J \beta_{jk, t} \phi_j (X_{ik}) ] + \varepsilon_{it}, \quad i = 1, \cdots, p; t=1,\cdots, T.
$$
The design matrix does not vary with $t$, neither does the projection matrix $\Pb$.

We make the following key assumptions:

\begin{assumption}
\label{a3}
\begin{description}

\item[(i) Pervasiveness:] With probability approaching one, all the eigenvalues of $\frac{1}{p}({\Pb}\Bb)'{\Pb}\Bb$ are bounded away from both zero and infinity as $p\to\infty$.

\item[(ii) Orthogonality:] $\mathbb{E}(u_{it}|X_{i1},...,X_{id})=0$, for all $i\leq p, t\leq T.$

\end{description}
\end{assumption}

The above  conditions require that the strengths of the  loading matrix should be as strong after the projection, and  $\Bb$  should be  associated with $\Xb$. Condition (ii) implies that if we apply $\Pb$ to both sides of $\Yb=\Bb\Fb'+\Ub$, then
$$
\Pb\Yb\approx\Pb\Bb\Fb',
$$
where $\Pb\Ub\approx 0$ due to the orthogonality condition. Hence the projection removes the noise in the factor model.
In addition, for the purpose of normalizations,  we assume  $\Cov(\bbf_t)=\Ib_K$, and that  $(\Pb\Bb)'\Pb\Bb$ is a diagonal matrix.

We now describe the rationale of the projected PCA. For simplicity, we ignore the effect of $\Pb\Ub$.
Let us consider the $p\times p$ covariance matrix of the projected data $\Pb\Yb$. The previous discussions show that
$
\frac{1}{T}\Pb\Yb(\Pb\Yb)'\approx  \Pb\Bb( \Pb\Bb)'.
$
Since $( \Pb\Bb)' \Pb\Bb$ is a diagonal matrix, the columns of $  \Pb\Bb$ are the eigenvectors of the $p\times p$ matrix $\frac{1}{T}\Pb\Yb(\Pb\Yb)'$, up to a factor $\sqrt{p}$. Next, consider the $T\times T$ matrix $
\frac{1}{T}(\Pb\Yb)'\Pb\Yb\approx\frac{1}{T}\Fb  ( \Pb\Bb)' (\Pb\Bb)\Fb'.
$
It implies
$$
\frac{1}{T}(\Pb\Yb)'\Pb\Yb\Fb\approx\Fb  ( \Pb\Bb)' (\Pb\Bb).
$$
Still by the diagonality of $( \Pb\Bb)' \Pb\Bb$, we infer that the columns of $\Fb$ are approximately  the eigenvectors of the $T\times T$ sample covariance matrix $
\frac{1}{T}(\Pb\Yb)'\Pb\Yb$, up to a factor $\sqrt{T}$. In addition, since the diagonal elements of $( \Pb\Bb)' \Pb\Bb$ grow fast as the dimensionality diverges, the corresponding eigenvalues are asymptotically the first $K$ leading eigenvalues of $\frac{1}{T}(\Pb\Yb)'\Pb\Yb$. This motivates the so-called ``projected PCA" (\cite{fan2014projected}), a new framework of estimating the parameters for factor analysis in the presence of a known space $\mathcal{X}$. The projected PCA can be more accurate than the usual PCA in the HDLSS scenario.  It applies really the PCA to the projected data (smoothed data) $\Pb\Yb$.

  Let $\widetilde \Vb$ be a $T\times K$ matrix, whose  columns  are the eigenvectors of the $T\times T$ matrix $\frac{1}{T}\Yb'\Pb\Yb$ corresponding to the larges $K$ eigenvalues.
 Following the previous discussions,   we respectively estimate the projected loading matrix $\Pb\Bb$ and   latent factors $\Fb$ by
$$\widetilde \Gb(\Xb)=\frac{1}{T}\Pb\Yb\widetilde\Fb, \quad \widetilde\Fb=\sqrt{T}\widetilde\Vb.$$

A nice feature of the projected-PCA is that the consistency is achieved even when the sample size $T$ is finite, as shown in \cite{fan2014projected}. Thus, it is particularly appealing in the HDLSS context. Intuitively, there are two sources of the approximation errors: (i)   $\Pb$ approximates $\Pb$ and  (ii) the normalized $\Bb$ approximates the leading eigenvectors of $\bSigma$. Neither of the approximation errors require a  large sample size $T$ in order to be asymptotically negligible. This implies the consistency under a finite $T$. See \cite{fan2014projected} for more detailed discussions on this aspect.

\subsection{Semi-parametric factor model}

In the model \eqref{eq4}, let  $\Gb(\Xb)$  and $\bGamma$ respectively denote the  $p\times K$ matrices of $g_k(\bX_i)$ and $\gamma_{ij}$. Then the matrix form of the model can be written as
$$
\Yb=[\Gb(\Xb)+\bGamma]\Fb'+\Ub.
$$
So the model assumes that the loading matrix can be decomposed into two parts: a part that can be  explained by $\Xb$ and the part cannot.   To deal with the curse of dimensionality, we assume $g_k(\cdot)$  to be additive:
$
g_k(\bX_i)=\sum_{l=1}^d g_{kl}(X_{il}) $, with $ d=\dim(\bX_i). $

Applying the projected-PCA onto the semi-parametric factor model,   \cite{fan2014projected}  showed that as $p,J \to\infty$, $T$ may either grow or stay constant,
$$
\frac{1}{\sqrt{T}}\|\widetilde \Fb-\Fb\|_2=O_P(\frac{1}{p}),\quad \frac{1}{\sqrt{p}}\|\widetilde \Gb(\Xb)-\Gb(\Xb)\|_2=O_P(\frac{1}{(p\min\{T, p\})^{1/2-1/(2\kappa)}}),
$$
where $\kappa$ is the degree of smoothness constant for $g_k(\cdot)$. Clearly  under the high dimensionality, the rate of convergence is fast even if $T$ is finite.  We refer the readers to \cite{fan2014projected} for more detailed discussions on the impacts of improved rates of convergence in factor models.


\section{Discussions}


This paper introduces several recent developments on estimating large covariance and precision matrices. We focus on two general approaches: rank-based method and factor  model based method.  We also extend the usual factor model  to a projected PCA setup, and show that the newly introduced projected PCA   is appealing in the high-dimensional-low-sample-size scenario. Such an approach has drawn growing attentions in  the recent literature on high-dimensional PCA (e.g., \cite{jung2009pca, shen2013consistency, SSZM, ahn2007high}).    In addition, we introduce the rank-based approaches, including the EPIC and  EC2  estimators, for estimating large precision and covariance matrices under the elliptical distribution family.  These rank-based methods are robust to heavy-tailed data and achieve the nearly optimal rates of convergence in terms of spectral norm errors.

A promising future direction is to combine the factor based analysis and rank-based analysis into an integrated framework.   For instance, consider  the  factor model
$$
    \bY_t=\Bb \bbf_t+{\bu_t}
$$
with observed factors $\{\bbf_t\}$. Here the idiosyncratic components $\bu_t$'s are heavy-tailed but follow the elliptical distribution.
Define the population Kendall's tau correlation between $u_{jt}$ and $u_{kt}$ as
\begin{align*}
\tau_{u,kj} = \PP\left((u_{jt}-\tilde{u}_{jt})(u_{kt}-\tilde{u}_k)>0\right) -\PP\left((u_{jt}-\tilde{u}_j)(u_{kt}-\tilde{u}_k)<0\right),
\end{align*}
where $\tilde{u}_j$ and $\tilde{u}_k$ are independent copies of $u_{jt}$ and $u_{kt}$ respectively. Let $\Rb_u$ be the correlation matrix of $\bu_t$, and $\Db_u$ be the diagonal matrix of the individual standard deviations of $\{u_{jt}\}$. Then $\bSigma_u=\Db_u\Rb_u\Db_u$.   For elliptical distributions,  we have
\begin{align}\label{e45}
\Rb_u = [\Rb_{u,kj}] = \left[\sin\left(\frac{\pi}{2}\tau_{u,kj}\right)\right].
\end{align}
Under the conditional sparsity condition, $\Rb_u$ is a sparse matrix.

Given the ``estimated residuals" $\{\widehat u_{it}\}$,  the sample version Kendall's tau statistic  is
\begin{align*}
\hat{\tau}_{u, kj} = \frac{2}{T(T-1)} \sum_{t<t'}\sign\Big((\widehat u_{kt} - \widehat u_{kt'})(\widehat u_{jt} - \widehat u_{jt'})\Big)
\end{align*}
for all $k \neq j$, and $\hat{\tau}_{u, kj}=1$ otherwise. We can   plug $\hat{\tau}_{u, kj}$ into  \eqref{e45} and obtain a  rank-based error correlation estimator
$
\hat{\Rb}_u = [\hat{\Rb}_{u, kj}] = \left[\sin\left(\frac{\pi}{2}\hat{\tau}_{u, kj}\right)\right].
$
We then apply thresholding on $\hat{\Rb}_u$  to produce a sparse matrix estimator:
$$
\widehat\Rb_u^{\mathcal{T}}=(\widehat\Rb^{\mathcal{T}}_{u,ij})_{p\times p},\hspace{1em}\widehat\Rb^{\mathcal{T}}_{u,ij}=\begin{cases}
1, & i=j;\\
h(\hat{\Rb}_{u, kj}; \omega_{T}), & i\neq j.
\end{cases}
$$
Here $h(.;\omega_{T})$ is a general thresholding rule as described in Section 2, with a properly chosen threshold value $\omega_T$.  The entry-dependent threshold can also  be  used.  Alternatively, we can apply the nearest positive definite projection to produce a sparse covariance estimator based on $\hat{\Rb}_u $.

Given the  estimated residuals,  standard deviations in $\Db_u$ can be estimated similarly as before. Specifically,  let  $\hat{m}_j$ be the estimators of   $\EE u_{jt}^2$ by solving:
\begin{align}
 \sum_{t=1}^T\psi\left((\widehat u_{jt}^2-{m}_{j})\sqrt{\frac{2}{T K_{\max}}}\right) = 0,\label{catoni-equation2}
\end{align}
where $K_{\max}$ is an upper bound of   $\max_j\Var(u_{jt}^2)$. Then the rank-based estimator of $\Db_u$ is a diagonal matrix $\widehat \Db_u$, whose diagonal  elements are   $
\hat{\sigma}_{u, j} = \sqrt{\max\bigl\{\hat{m}_j, K_{\min}\bigr\}}$,
where $K_{\min}$ is a lower bound of $\min_j\EE u_{jt}^2$ and is assumed to be known. This leads to the rank-based error covariance estimator:
$$
\widehat\bSigma_u=\widehat \Db_u \widehat\Rb_u^{\mathcal{T}} \widehat \Db_u.
$$

When the factors are observable,  the residuals should be obtained by estimating $\Bb$. The robust regression estimator $\widehat \Bb$ can be employed, e.g., $L_1$ regression. With the estimated $\Bb$, we set  $\widehat \bu_t= \bY_t-\widehat \Bb \bbf_t$.
 The final factor-based covariance estimator is then given by:
$$
\widehat\bSigma=\widehat\Bb\widehat\Cov(\bbf_t)\widehat\Bb'+\widehat\bSigma_u.
$$
 The resulting estimator is expected to naturally handle heavy-tailed data.

When the common factors are latent,  they need to be estimated using robust PCA (that is, applying PCA on the rank covariance matrix of $\bY_t$).
  The theoretical properties of such hybrid estimators are left for future investigations.



\bibliographystyle{ims}
\bibliography{graphbib}

\end{document}